\documentclass[12pt,preprint]{aastex}
\usepackage[left]{lineno}
\usepackage{subfigure}
\usepackage{braket}

 \slugcomment{}

\shorttitle{}
\shortauthors{}

\begin{document}


\title{Probing Rotation of Core-collapse Supernova
    with Concurrent Analysis of Gravitational Waves and Neutrinos}


\author{
Takaaki Yokozawa\altaffilmark{1}, 
Mitsuhiro Asano\altaffilmark{1}, 
Tsubasa Kayano\altaffilmark{2},
Yudai Suwa\altaffilmark{3, 4},\\
Nobuyuki Kanda\altaffilmark{1},
Yusuke Koshio\altaffilmark{2}, 
and Mark R. Vagins\altaffilmark{5, 6}    
}


\altaffiltext{1}{Graduate School of Science, Osaka City University, Sumiyoshi-ku, Osaka 558-8585, Japan}
\altaffiltext{2}{Department of Physics, Okayama University, Okayama, Okayama, 700-8530, Japan}
\altaffiltext{3}{Yukawa Institute for Theoretical Physics, Kyoto University, Oiwake-cho, Kitashirakawa, Sakyo-ku, Kyoto 606-8502, Japan}
\altaffiltext{4}{Max-Planck-Institut f\"ur Astrophysik, Karl-Schwarzschild-Str. 1, 85748 Garching, Germany}
\altaffiltext{5}{Kavli Institute for the Physics and Mathematics of the 
Universe (WPI), University of Tokyo Institutes for Advanced Study, 
University of Tokyo, Kashiwa, Chiba 277-8583, Japan} 
\altaffiltext{6}{Department of Physics and Astronomy, University of California, Irvine, Irvine, CA 92697-4575, USA}


\begin{abstract}

The next time a core-collapse supernova (SN) explodes in our galaxy, 
various detectors will be ready and waiting to detect its emissions of  
gravitational waves (GWs) and neutrinos. 
Current numerical simulations have successfully introduced 
multi-dimensional effects to produce exploding SN models, 
but thus far the explosion mechanism is not well understood. 
In this paper, we focus on an investigation of progenitor core 
rotation via comparison of the start time of GW emission and 
that of the neutronization burst.
The GW and neutrino detectors are assumed to be, respectively, the 
KAGRA detector and 
a co-located gadolinium-loaded water Cherenkov detector, 
either EGADS or GADZOOKS!.
Our detection simulation studies show that for 
a nearby supernova (0.2~kpc) we can confirm the lack of 
core rotation close to 100$\%$ of the time, and the presence of 
core rotation about 90$\%$ of the time.  
 Using this approach there 
is also potential to confirm rotation for considerably more 
distant Milky Way supernova explosions.

\end{abstract}





\section{Introduction}\label{sec/intro}

Supernova explosions are among the most energetic events in the
universe. Core collapse supernovae are the final phase in the
evolution of massive stars with $M\gtrsim 8M_\odot$. Most of the
released gravitational energy, $\sim 10^{53}$ ergs, is emitted as
neutrinos and only a small fraction ($\sim 1\%$, so $\sim10^{51}$
ergs) is used to produce the violent explosion
itself \citep{beth90}. Although the historic detection of neutrinos
from SN 1987A \citep{kam1987a, imb1987a, baksan_1987a} validated our
basic picture of core collapse (e.g. the total energy, the emission
timescale of neutrinos, etc.), there are still large uncertainties
concerning the explosion mechanism itself. In addition to photons and
neutrino observations, the next nearby (probably galactic) supernova
will also likely be observed using a relative newcomer to
multi-messenger astronomy, namely gravitational waves (GWs).  Owing to
their weak coupling with matter they have the potential to provide us
with information about the innermost part of these violent phenomena,
and hence could serve as a unique window into the explosion
mechanism \citep[see][and references therein]{ott09,kota12}.
The new multi-messenger astronomy era will include observation of
broad-band photons (from radio to gamma), multi-energy neutrinos (from
MeV to PeV), and multi-frequency GWs (from Hz to kHz) using various
telescopes and detectors already in operation as well as those coming
online in the near future. Using these signals we can see different
aspects of astronomical objects. What's more, after decades of effort
recent developments in the numerical simulation of core-collapse
supernovae are making remarkable progress: there are several
simulation studies showing successful shock expansion (see
Sec. \ref{sec/snmodel}.). In light of these facts, we will give some
prospects of what we can learn from the next nearby core-collapse
supernova,  
with special focus on the presence of progenitor core rotation.


We employ theoretical predictions of self-consistent signals of
neutrinos and GWs based on recent simulations of multi-dimensional
neutrino-radiation hydrodynamics \citep[e.g.,][]{suwa10}. By
considering a realistic detector simulation that consists of not only
responses for both GWs and neutrinos but also takes into account noise
and statistical behavior, we study the feasibility of extracting
information concerning the most crucial period in the death of massive
stars. We define characteristic times: epoch time of GW ${\cal
T}_{GW}$ and of neutronization burst ${\cal T}_{\nu}$. Supposing that
the core rotates rapidly, the GW would be detectable prior to
neutronization burst (i.e. ${\cal T}_{GW}<{\cal T}_{\nu}$ ) and vice
versa \citep[see][for time determination of core bounce by neutrinos
alone]{pagl09,halz09}. 
%
This work is complementary to previous studies, in which a
principal component analysis or Baysien analysis and many other
approaches have been employed to extract infomation of core rotation
from GW signals around core bounce alone \citep[][and many
others]{hayamasn, rotation1,rotation2, rotation3, rotation4,
rotation5, rotation6, rotation7}. Since GW waveforms are significantly
affected by the various initial conditions of simulations and status
of GW detector, complex analyses are required if we use only
GW. 
Instead, our method tackles this issue from a multi-messenger astronomical perspective; we simply compare the GW epoch and neutronization burst times. Our analysis has the potential to provide not only evidence of progenitor core rotation, but also the time of core bounce. By combining them, we can expect a more robust extraction of the rotation law and velocity of rotation.


For our reference GW detector we use KAGRA, which is currently under
construction \citep{KAGRA_detector}, and for our neutrino detector
use either EGADS, which is now operating \citep{EGADS_proc}, 
and GADZOOKS! \citep{gadzooks}, an upgraded version of Super-Kamiokande 
expected to start taking data in 2017.
This is because (i)
the supernova neutronization burst generates copious amount of electron-type
neutrinos, 
making these gadolinium-loaded water Cherenkov detectors 
more suitable than the classic Super-Kamiokande configuration filled with 
with ultrapure water\footnote{Here we neglect neutrino oscillation
effects for simplicity.}\citep{snsk}, and (ii) since KAGRA, EGADS, and
GADZOOKS! are all located in the same place, i.e. the Kamioka mine, 
signal travel time between detectors does not have to be
taken into account.

Section \ref{sec/snmodel} describes the numerical simulation and
supernova signals which are used in this analysis.
Section \ref{sec/detectors} describes the characteristics of KAGRA, EGADS, 
and GADZOOKS!.  Section \ref{sec/result} shows the method of time
extraction of ${\cal T}_{GW}$ and ${\cal T}_{\nu}$, with the result of 
determining an estimator of progenitor core rotation. Finally,
Section \ref{sec/summary} contains a summary and discussion of
our results.

\section{Numerical Simulation of Core-collapse Supernova}\label{sec/snmodel}
In this section, we illustrate the numerical simulation of
core-collapse supernova explosions. First, we shortly explain the
basics of simulations. Secondly, numerical methods used in this study
are given. Thirdly, hydrodynamic features are expressed, and finally
neutrino and gravitational wave emissions are presented.

\subsection{Basics}

Since there are various types of physics involved in core-collapse
supernova explosions, detailed numerical simulations are
indispensable. For instance, gravity describes how the matter
collapses and how much energy is released during the collapse. In
addition, a final outcome of supernovae are neutron stars (NSs), so
that the nuclear force, which determines the structure of a NS, is
critically important. Neutrino interaction rates, which give the
cooling rates of the core as well as heating rates of post-shock
material by neutrino absorption (this will be explained later), are
treated in great detail to give a quantitatively correct answer.  This
is because the total amount of neutrino emission is $\sim 10^{53}$
erg, while the explosion energy itself is $\sim 10^{51}$ erg,
therefore only a few percent of energy deposition by neutrinos can
drive the explosion; indeed, this is the so-called standard scenario
of core-collapse supernovae (see \citealt{colg66,beth85} for original
idea and \citealt{kota12,jank12,burr13} for recent reviews).

Due to the complexity of the system, we have not achieved fully
consistent explosions using simulation thus far. However, several
exploding simulations arising from the neutrino heating mechanism have
been reported in the last decade 
  (\citealt{bura06,mare09,suwa10,muel12b,brue13,pan15} for 2D and
  \citealt{taki12,mels15,lent15,muel15} for 3D).\footnote{Note that
  some simulations of multi-dimensional neutrino radiation
  hydrodynamics also reported the failure of the explosion by neutrino
  heating \citep{burr06,ott08,hank13,dole15}, so that this mechanism
  still contains ambiguities.} These simulations exhibited smaller
explosion energy, i.e., $\sim 10^{49}$--$10^{50}$ ergs, than the
observationally required value, $\sim 10^{51}$ ergs, leading to
continuous accretion onto a NS beyond the maximum supportable mass,
inevitably resulting in an eventual collapse to a black hole instead
of a NS.  More recently, although the small explosion energy problem
remains, there has been a {\it successful} exploding simulation, which
means the mass accretion onto a NS ceases and the final mass
eventually settles into $\sim 1.3 M_\odot$, by \cite{suwa14a} using a
progenitor with 11.2 $M_\odot$ at zero-age main sequence phase. This
explosion is a consequence of a steep density gradient between the Si
and O layers, which results in a rapid decrease of the mass accretion
rate onto the shock wave \citep{suwa14b}. After this shell passes
through the shock, the shock begins to expand and the system
eventually produces the explosion. In this study, we use the same
progenitor and the same simulation code.

\subsection{Methods}

The numerical methods used in this study are the same as
\citet{suwa10,suwa11b,suwa13b,suwa14b,suwa14a}. In this code, we solve
neutrino-radiation hydrodynamics equations, which consist of
hydrodynamics equations and neutrino radiative transfer equations (see
\citealt{suwa13b} for more details).

In the hydrodynamics simulations we employ axial symmetry and
Newtonian gravity.  The neutrino transfer equations for $\nu_e$ and
$\bar\nu_e$ are solved with an isotropic diffusion source
approximation (IDSA) by \cite{lieb09} and heavier leptonic
  neutrinos are not taken into account. The weak interaction rates
are based on description of \cite{brue85}. In the current study, we
neglect neutrino-electron scattering.  The nuclear equation of state
employed is \cite{latt91} with an incompressibility parameter of
$K=220$ MeV. As for the initial condition, we employ $11.2 M_\odot$
model from \citet{woos02}, which was used in number of previous works.

The rotation is imposed via a shellular rotation law as
\citep[e.g.][]{erig85}
\begin{equation}
\Omega(r)=\Omega_0\frac{r_0^2}{r^2+r_0^2},
\end{equation}
where $\Omega(r)$ is an angular velocity with $r$ being the radius
from the center, $\Omega_0$ is the angular velocity at the center, and
$r_0$ is a radius that determines the degree of differential
rotation. Here, we employ $r_0=1000$ km. In order to investigate how
the rotation affects emissions of gravitational waves and neutrinos,
we perform simulations with two different rotation strengths of
$\Omega_0=$ 0.0$\pi$ and 1.0$\pi$ radian s$^{-1}$.  
%

  Since little is known about the rotation rates of central cores 
  just prior to core collapse, we just employ these values.
  The latter one
  (period $P_0=2\pi/\Omega_0=2$ s) is relatively faster than the
  current estimation based on stellar evolutionary calculations
  ($P_0\sim 100$ s), which take into account angular momentum transfer
  processes \citep[e.g.,][]{hege05}. Note that these calculations
  contain many ambiguities for angular momentum transfer. 
  The direct observational constraints for stellar core rotation have
  been given only for low-mass red giant stars using asteroseismology
  \citep{beck12} and the core rotation rates of massive stars making
  supernovae are observationally still uncertain.
  The spin period distributions of young pulsars also 
  give constraints on rotation of precollapse cores, 
  which imply the core rotation periods prior to collapse are 
  longer than several tens of seconds, 
  if the angular momentum of the inner cores are conserved
  \citep[e.g.,][]{ott06}. On the other hand, subsequent GW emission
  induced by r-mode instability of NS may slow down spin of rapidly
  rotating (with almost breakup velocity) pulsars to period of
  $\sim$10 ms, which is compatible with estimated birth periods of
  rapidly rotating pulsars like PSR J0537-6910
  ($P_\mathrm{pulsar}=16.1$ ms) and the Crab pulsar
  ($P_\mathrm{pulsar}=33.5$ ms) \citep{lind98}. 
  Therefore, at present the core rotation profiles 
  during the precollapse phase are highly uncertain, 
  and the observation of GW can be a smoking gun of this important issue.

In this paper, we evaluate the gravitational wave emission from
aspherical motion of fluids via the Newtonian quadrupole formulas of
\cite{moen91}.  In the axisymmetric case, the components of the
dimensionless gravitational wave strain in the transverse-traceless
(TT) gauge are represented as
\begin{equation}
h^{TT}_{\theta\theta}=-h^{TT}_{\phi\phi}=\frac{1}{8}\sqrt\frac{15}{\pi}\sin^2\alpha\frac{A^{E2}_{20}}{D}\equiv h_+,
\end{equation}
where $\alpha$ is the angle between the symmetric axis and the line of
sight of the observer. In this study, we assume $\sin\alpha=1$ for
simplicity. Due to the axial symmetry, the other component,
$h_\times$, is vanishing. $A^{E2}_{20}$ is a coefficient of the mass
quadrupole contribution, which can be expressed by hydrodynamic
variables as
\begin{eqnarray}
A^{E2}_{20}&=&\frac{16\pi^{3/2}}{\sqrt{15}}\frac{G}{c^4}
\int^{1}_{-1}d\mu
\int^{\infty}_{0}dr~
r^2\rho
\left[v_r^2(3\mu^2-1)+v_\theta^2(2-3\mu^2)-v_\phi^2\right.\nonumber\\
&&~\left.-6v_r v_\theta \mu\sqrt{1-\mu^2}-r\partial_r\Phi(3\mu^2-1)+3\partial_\theta\Phi\mu\sqrt{1-\mu^2}
\right],
\end{eqnarray}
where $G$ is the gravitational constant, $c$ is the speed of light,
$\mu=\cos\theta$, $\rho$ is the density, $v_r$, $v_\theta$ and
$v_\phi$ are velocity vector components in $r$, $\theta$, and $\phi$
directions, and $\Phi$ is the gravitational potential. $\partial_r$
and $\partial_\theta$ are $\partial/\partial r$ and $\partial/\partial
\theta$, respectively. $D$ it the distance between the observer and
the source.
We do not discuss the gravitational wave emission from anisotropic
neutrino emission \citep{epst78,burr96,muel97} because this component
contributes to GW at later time and does not affect GW around the
bounce time.

\subsection{Hydrodynamic Features}

After the simulation sets in, the gravitational collapse begins and
the central density increases. Around 180 ms, the central density
reaches the nuclear density ($\sim 3\times 10^{14}$ g cm$^{-3}$),
which indicates the formation of a protoneutron star (PNS). The
equation of state becomes much stiffer above this density so that the
infall of material ceases and a shock wave is produced at the surface
of the PNS.
\footnote{The inner core mass at the bounce in our simulation is $\sim
  0.7M_\odot$. This is larger than the slowly rotating model in
  \cite{dfm}, in which they showed that the inner core masses for
  slowly rotating models range between $0.4$--$0.5 M_\odot$, depending
  on the progenitor models and equation of state. The difference is
  from omission of electron scattering, which lowers the electron
  fraction and leads to smaller effective inner core mass, and
  neglecting general relativistic effects in this study.}  The shock
propagates outward initially, but loses its energy due to the
photodissociation of iron and neutrino emission so that it
decelerates. About 100 ms after the bounce, which is defined here by
the time of the maximum central density, the shock expands again owing
to the neutrino heating mechanism. All simulations used in this study
result in explosion, i.e., we observe the shock expansion above a few
thousands km. As for the nonrotating model, \cite{suwa13b} and
\cite{suwa14a} give more information, which are valid even for
rotating models of this paper, since the rotation is not too strong to
change the whole picture.

\subsection{Neutrinos and Gravitational Waves}

\begin{figure}[ht]
 \centering
 \subfigure[$L_{\nu_e}$]{\includegraphics[width=0.45\textwidth]{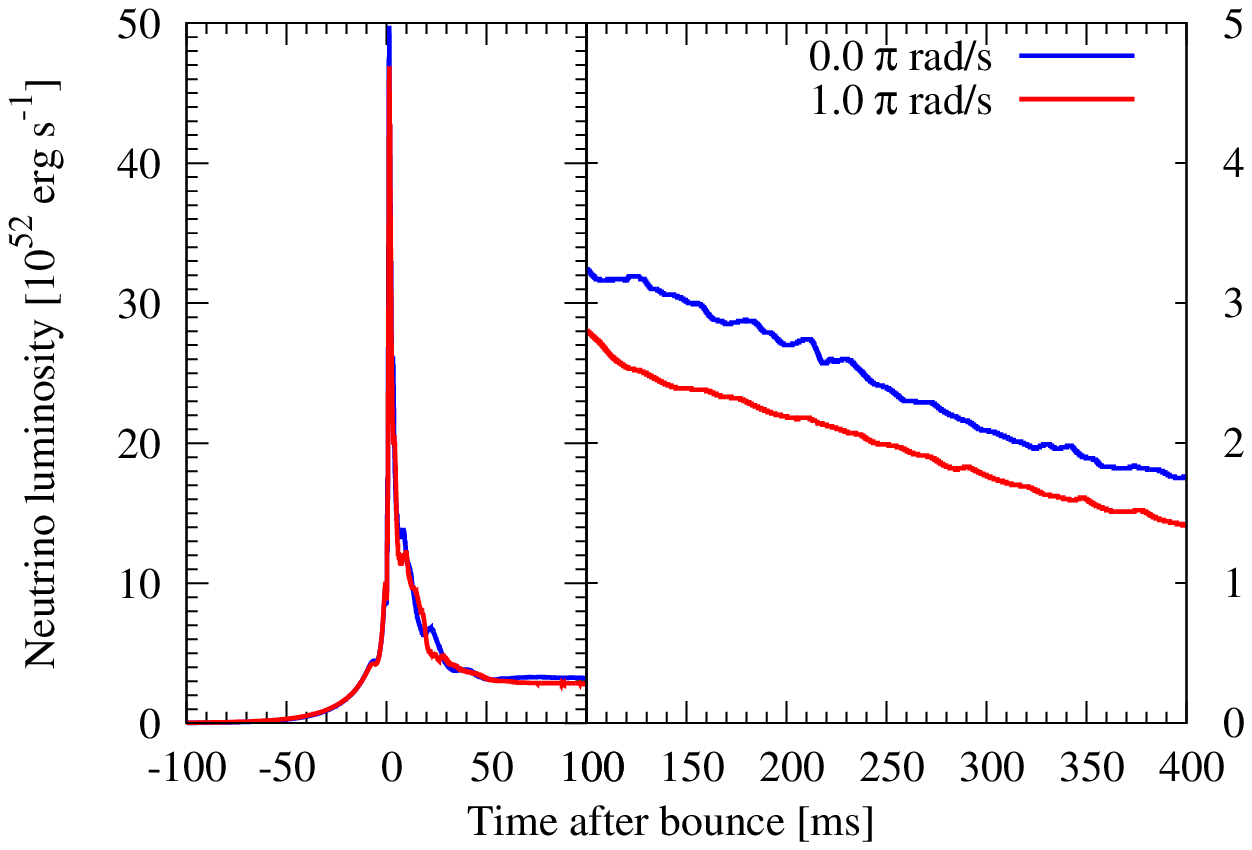}}
 \subfigure[$L_{\bar\nu_e}$]{\includegraphics[width=0.45\textwidth]{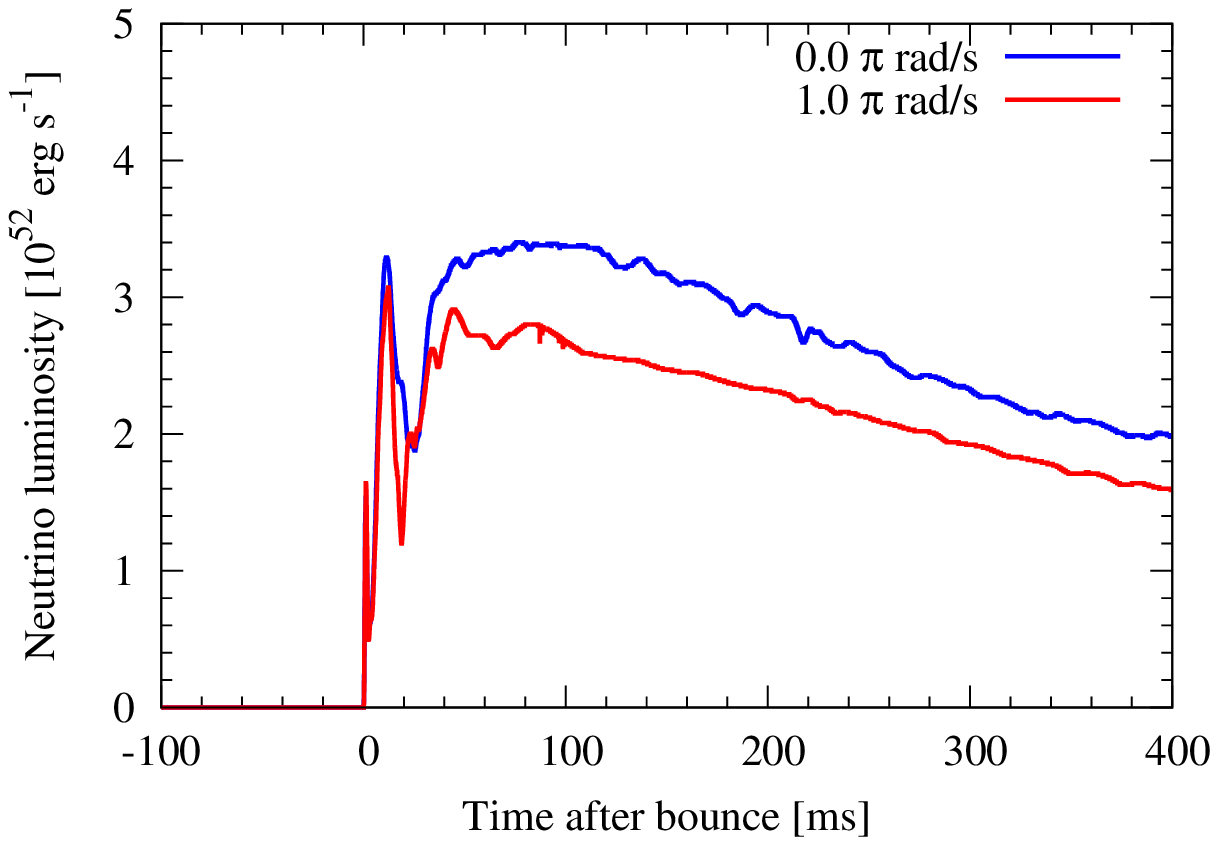}}
 \subfigure[$\braket{E_{\nu}}$]{\includegraphics[width=0.45\textwidth]{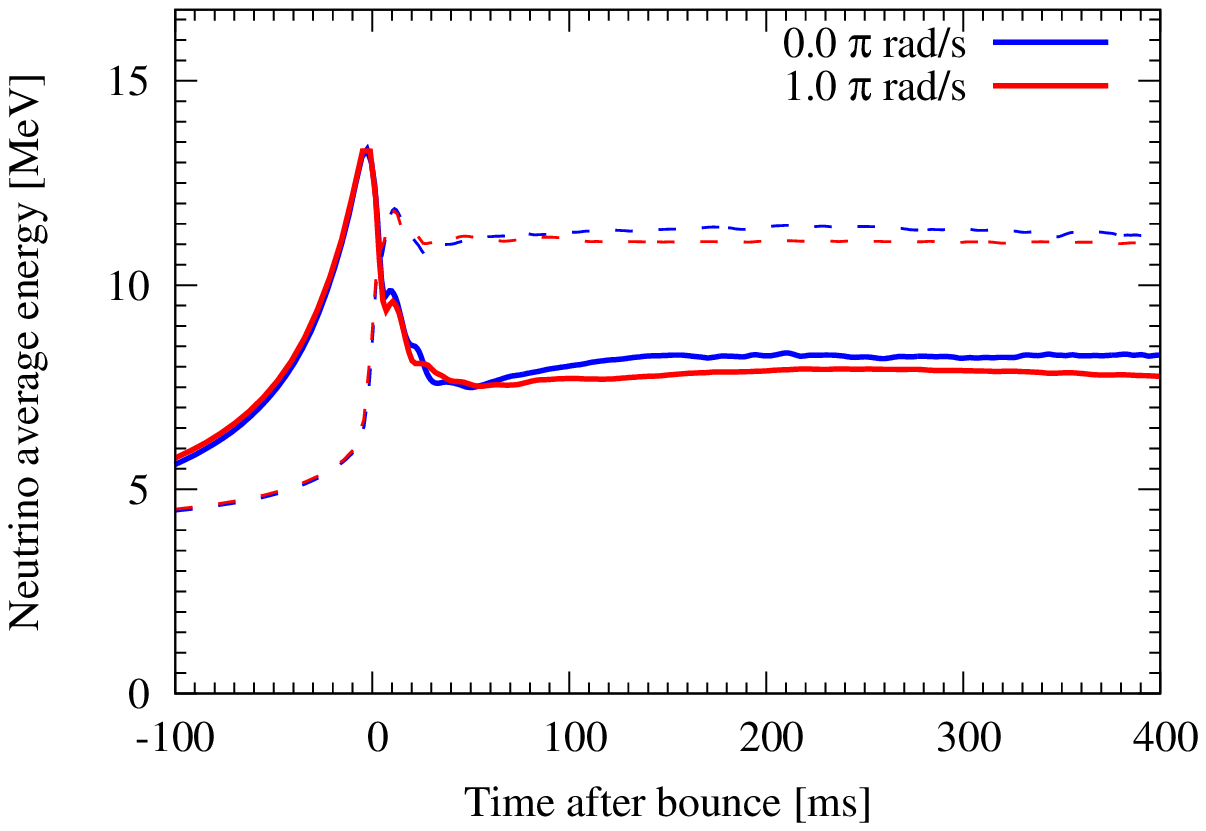}}
 \caption{Time evolution of luminosities of electron-type neutrinos
   ($\nu_e$; panel(a)) and antineutrinos ($\bar\nu_e$; panel (b)) and
   average energy of them (panel (c)). Each line represents models
   with different initial rotation rates, i.e., no
   rotation($\Omega_0=0.0\pi$) (blue) and 
   strong rotation $\Omega_0=1.0\pi$ rad s$^{-1}$ (red).
   The solid and dashed lines in panel (c) show $\braket{E_{\nu_e}}$
   and $\braket{E_{\bar\nu_e}}$, respectively. Note that left and
   right sides in panel (a) have different scales.}
 \label{2-neutrino}
\end{figure}

Figure \ref{2-neutrino} represents the time evolution of luminosities
of electron-type neutrinos ($\nu_e$; panel (a)) and antineutrinos
($\bar\nu_e$; panel (b)), and their average energies.  One can see the
general trend does not depend on the initial rotation rate very
much. In the first few tens of ms for $\nu_e$, there is the so-called
neutronization burst, which is generated by the rapid absorption of
electrons by protons (electron capture; $\mathrm{e}+\mathrm{p}\to
\mathrm{n}+\nu_e$) in the regime between the shock and PNS, causing
emission of large amounts of $\nu_e$. After that the neutrino
luminosity is gradually decreasing, but a large luminosity is still
observed. On the other hand, $\bar\nu_e$ does not have such a spiky
structure due to lack of positrons before bounce. Both $\nu_e$ and
$\bar\nu_e$ exhibit that the rotating model ($\Omega_0=1.0\pi$ rad
s$^{-1}$) results in slightly smaller luminosity due to stronger
centrifugal force and slower contraction of the PNS.
Therefore, we can argue that the currently employed rotation strength
does not significantly change the generic picture about neutrinos.
Note that we calculated neutrino luminosity at the outer boundary
based on streaming particles of IDSA, a stationary-state solution of
Boltzmann equation, so that the propagation time between emission site
and the outer boundary is neglected. This approximation is fully
consistent with the estimation of gravitational waves using the
quadrupole formula.

\begin{figure}[ht]
 \centering
 \includegraphics[width=0.8\textwidth]{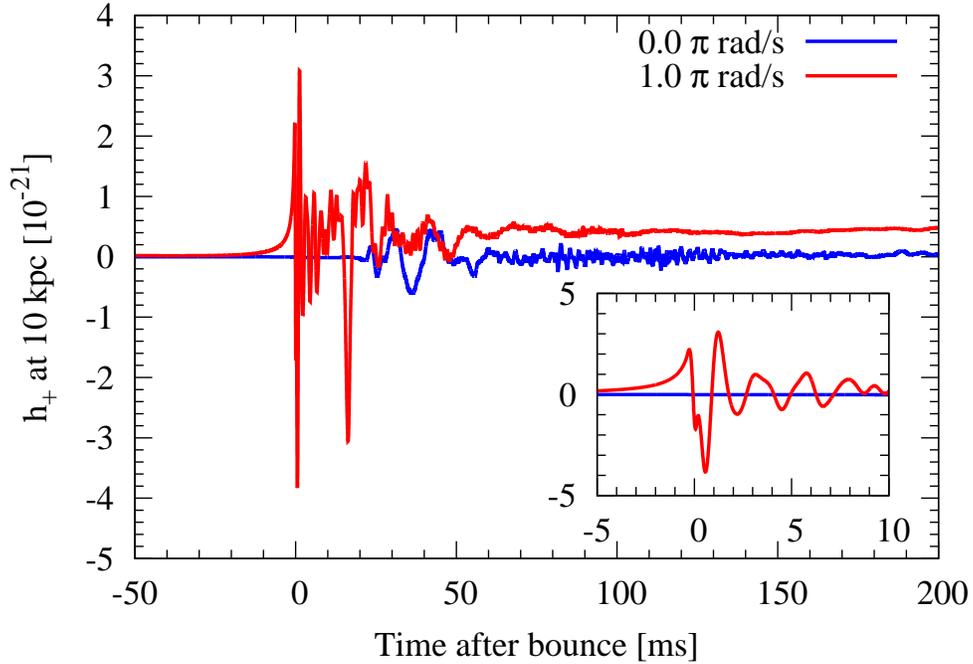}
 \caption{GW amplitude as a function of time for a core-collapse
   supernova occurring 10 kpc from the observer. Each line represents
   models with different initial rotation rate, i.e., no rotation
   ($\Omega_0=0.0\pi$)
   (blue) 
   and strong rotation$\Omega_0=1.0\pi$ rad s$^{-1}$ (red).
   The small panel shows features around the bounce
   time. The fast rotation model exhibits a large amplitude GW at
   the bounce, while the slow rotation models have very
   small amplitudes. Since the later phase (i.e., a few tens of ms after
   the bounce) activity is dominated by convection motion, all
   models imply similar amplitude in the later phase.}
 \label{2-gw}
\end{figure}

Figure \ref{2-gw} depicts the time evolution of the gravitational wave
signal. The gravitational wave signals obtained in this study are all
classified as {\it type-I} \citep{zwer97,dfm}, which have a strong
peak at the core bounce when the central density exceeds the nuclear
density.  Note that in the early phase the gravitational wave strength
strongly depends on the initial rotation (see small panel). As for
model without rotation,
the density structure is almost spherically symmetric so that there is
no
GW emission. On the other hand, the strongly rotating model exhibits
strong GW emission at the time of bounce because the centrifugal force
makes the core asymmetric. Therefore, we can constrain the rotation
strength by detecting GW at just the time of bounce. We can constrain
the bounce time in turn using neutrino data as shown in
Fig. \ref{2-neutrino}, in which we show that the neutrino emission
does not depend on the rotation strength so that neutrinos are a
guaranteed signal from core-collapse supernova.


\section{Detectors}\label{sec/detectors}
 In this section, we introduce the KAGRA, EGADS, and GADZOOKS! detectors,
which are used to determine progenitor core rotation in this paper.

As described in the introduction, 
one of the important benefits of employing 
is the close proximity of these three detectors. 
This makes it possible to avoid any significant time-lag in arrival 
times of gravitational wave and neutrino signals  
due to distance between the detectors.
In gravitational wave detection, 
the angular resolution of the source direction is 
not as good as that of optical telescopes, 
even in multiple-detector observations. 
According to the diffraction limit of wavelength 
($\sim$300 km for 1 kHz) and interval of detectors ($\sim$1000 km), 
typical angular resolution for the kHz band burst wave is 
only a few degrees \citep{sndirgw}. 
Water Cherenkov detectors have a
similar, few-degree angular resolution for supernova explosions  
$\sim$ 10 kpc away (i.e., near the galactic center) \citep{sndirnu1, sndirnu2}. 
These facts mean that the correction of arrival time 
based on direction for well-separated detectors will have a larger error than 
the timing accuracy which is 
required in this paper study, $<1$ msec,  
for neutrino and gravitational wave signals only. 
Closely placed detectors do not suffer from this problem.

\subsection{KAGRA Detector}

KAGRA is a laser interferometric gravitational wave detector which is 
being constructed in the Kamioka mine in Gifu, Japan \citep{bib:kagra_design}. 
The KAGRA interferometer consists of two 3-kilometer long laser cavity arms;  
all optical and vacuum systems are located inside a mountain, providing  
a silent and stable environment. KAGRA will employ cryogenic mirrors made 
of monolithic crystals of sapphire to reduce the thermal noise.
Its target sensitivity is a few $\times 10^{-24}$ ${\rm [1/\sqrt{Hz}]}$ in 
strain `$h$' of the space-time metric around 100 Hz \citep{KAGRA_detector}.
\footnote{\cite{bib:kagra_sensitivity}} There are other gravitational wave 
detectors which will have similar sensitivity, 
i.e. LIGO in the US \citep{bib:LIGO} 
and Virgo in Europe \citep{bib:Virgo}. Their upgraded configurations 
\footnote{aLIGO ; https://www.advancedligo.mit.edu/} 
\footnote{aVirgo ; https://wwwcascina.virgo.infn.it/advirgo/} and 
a global observation network of four gravitational wave detectors are expected 
to be in operation in late 2017 or 2018.


To achieve these detectors' high sensitivity various things are required: a 
long baseline to integrate strain effect, a high-power laser to protect 
against quantum shot noise, high-quality and heavy mirrors, a vacuum system to 
eliminate air fluctuations, many advanced technique of optics including 
quantum optics, and so on. KAGRA employs additional special techniques, 
particularly in its underground site and cryogenic mirrors.

\subsection{EGADS and GADZOOKS! Detectors}
The EGADS (originally standing for Evaluating Gadolinium's Action on
Detector Systems) detector \citep{EGADS_proc} is a gadolinium (Gd)
loaded water Cherenkov detector initially built as a demonstrator for
GADZOOKS! \citep{gadzooks}, the proposal to load Super-Kamiokande (SK)
with a water-soluble gadolinium salt. EGADS is located in the Kamioka
mine which is in the same mountain as both SK and KAGRA.

GADZOOKS! envisions adding 0.2$\%$ by mass of gadolinium sulfate
(Gd$_2$(SO$_4$)$_3$) into SK in order to facilitate the efficient
detection of neutrons from Inverse Beta Decay (IBD) reactions
($\bar{\nu_e} + p\to e^+ + n$). Gadolinium has the highest capture
cross section for thermal neutrons of any naturally occurring
substance, and emits an 8.0 MeV gamma cascade following capture.  This
can be easily detected in water Cherenkov detectors like SK, making
neutrons visible and thereby tagging up to 90$\%$ of the IBD events as
genuine due to the coincidence in both time ($\sim$30 ms) and space
($\sim$1 meter) between the prompt positron and the delayed gadolinium
neutron capture cascade \citep{ntagging}.

The efficient neutron tagging made possible by gadolinium loading
brings many benefits to the venerable water Cherenkov technology 
in terms of supernova neutrino detection.
First of all, the distinctive ``gadolinium heartbeat'' - the double pulse of 
positron Cherenkov light and neutron capture gamma cascade - from IBD events 
in a Gd-loaded detector 
will instantly identify any galactic supernova as genuine 
\citep{nextsn}.   Next, by allowing event-by-event tagging of the 
copious IBD events, a pure $\bar{\nu_e}$ time structure and energy spectrum 
can be precisely characterized and then 
subtracted away, exposing other, more subtle signals.  
Doing so yields a variety of
powerful advantages, including: improving the determination of
the supernova's position in the sky, measuring the temperature 
of the burst, spotting the moment of birth of a black hole, 
and potentially identifying the early neutronization burst.

The EGADS detector consists of a cylindrical stainless steel tank, whose
height and radius are 6.7 m and 6.5 m, respectively.  A total of 240
inward-facing photodetectors line the inner walls of the EGADS tank.
Most of these ($\sim$90$\%$) are SK-style 50-cm diameter
photomultiplier tubes (PMTs), while the rest are prototype light
detectors of various sizes and designs being considered for use in the
future Hyper-Kamiokande project \citep{HK}.  The resulting active
light collecting surface area in EGADS is 40$\%$, the same as in SK.
The EGADS tank contains a total of 200 tons of Gd loaded water, about
100 tons of which is in front of the PMTs.

Following the R$\&$D phase of operations (2010-2014), EGADS has been
repurposed as a dedicated supernova neutrino detector, with the
acronym now standing for Employing Gadolinium to Autonomously Detect
Supernovae.  Deadtime-free front-end electronics \citep{qbee} have
been purchased, the same type used in SK since 2008, and will be
installed in EGADS in 2015.  From that point on the detector will run
continuously as a supernova neutrino detector with realtime online
event reconstruction.

In addition to lacking effective neutron tagging in its current pure-water
configuration, SK cannot record all
of the data produced by nearby supernovae \citep{snbetel}.  By
contrast, EGADS not only already has highly efficient tagging but also
full supernova sensitivity for the entire Milky Way galaxy; it would
expect to record about 40 events for a core-collapse explosion at the
galactic center and 100,000 events for an explosion at the distance of
Betelgeuse.  
If and when GADZOOKS! goes forward, SK will of course gain the 
strong advantages of efficient neutron tagging -- which will be even 
more powerful than in EGADS due to SK's much greater size -- but SK's  
DAQ limitations will persist for very close bursts.  For this reason, 
EGADS can be expected to continue to play a useful role 
as a supernova neutrino detector well into the future, whether or not 
gadolinium is added to Super-Kamiokande.


\section{Detection Simulation}\label{sec/result}
In this section we describe the detector signal simulation for KAGRA, EGADS,
and GADZOOKS!, and the method of how to determine progenitor core rotation
from these detectors.  The analysis path is as follows:
(i) Run supernova detection simulation for a given situation.
(ii) Extract the epoch time of GW, ${\cal T}_{GW}$, and neutronization
burst, ${\cal T}_{\nu}$.
(iii) Compare these times and determine the presence of rotation or its absence.
(iv) Loop 100,000 times and evaluate $P_r$, which is the probability 
of core rotation.
The initial angular momentum of progenitor core rotation in the
supernova models are given two different magnitudes: 
$\Omega_0=$ 0.0$\pi$ and 1.0$\pi$ rad s$^{-1}$. 
Four supernova scenarios are considered: 
uniform distributions of explosions at distances of 0.2 kpc or 1.0 kpc, 
for which the detection simulations of KAGRA and the EGADS detector are 
used; and explosions at the galactic center or distributed throughout 
the galaxy, for which the detection simulations of KAGRA and GADZOOKS! are used.
This distinction is made is because of serious doubts (as discussed in the 
EGADS/GADZOOKS! section) regarding the current SK DAQ system's performance 
at the extremely high rates expected for nearby explosions.
The galactic center is defined as (right ascension, declination) 
= (17h42m26.603s, -28$^{\circ}$55'00.445'') and its distance is set to 10 kpc.
The galactic distribution is generated by an exponential distribution 
model \citep{galaxydist}, where the differential number of supernovae, dN, 
is proportional to
\begin{equation}
 dn \propto R \cdot dR \cdot dz \cdot e^{-\frac{R^2}{2R^2_0}} \cdot e^{-\frac{|z|}{h}},
\end{equation}
where $R_0$ and h are variation factors and
\begin{eqnarray}
 R = 3.5 \mathrm{kpc}\\ \nonumber
 h = 0.32 \mathrm{kpc}.
\end{eqnarray}
Note that the incident direction strongly affects GW detector response,
but has little bearing on neutrino detection.

\subsection{GW Analysis}

\subsubsection{Detector Signal}

The output signal of the interferometric detector, $s(t_i)$, will be
sampled with finite frequency 1/$\delta t$, and can be written as
\begin{equation}
s(t_i) = h(t_i) + n(t_i),
\end{equation}
where $i$=0,1,2,.. is sample index and $h(t_i)$ is gravitational
signal from supernova with beam pattern, polarization of GW emission,
and distance from the sources taken into account. The 16,384 Hz
sampling frequency of the KAGRA detector is used. $n(t_i)$ is time
domain of detector noise.

\begin{figure}[ht]
 \centering
 \includegraphics[width=0.7\textwidth]{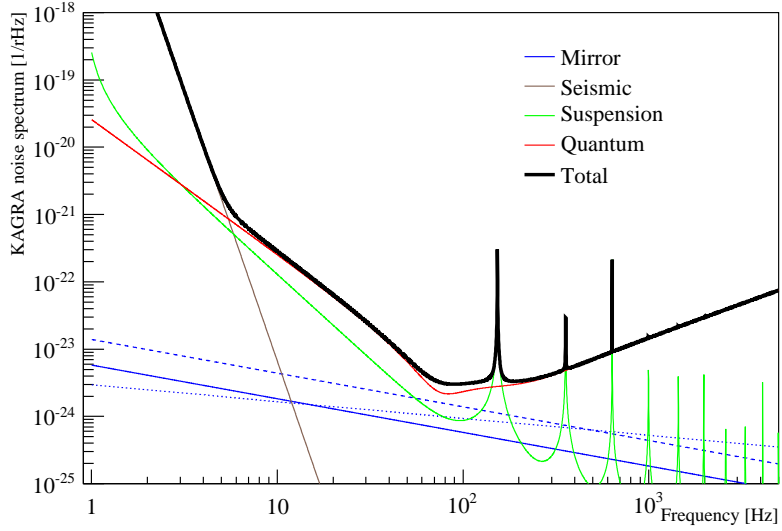}
 \caption{KAGRA official sensitivity curve. These curves are estimated 
          from an incoherent sum of the fundamental noise sources. 
          The colors show each type of environmental noise and the black line shows 
          total noise. Blue lines are mirror related noise, brown shows seismic noise,
          green shows suspension thermal noise, and red shows quantum noise, respectively.}
 \label{fig:kagranoise}
\end{figure}

When generating $n(t_i)$, we assume stationary and Gaussian detector
noise of the KAGRA detector.
Figure \ref{fig:kagranoise} shows the sensitivity curve of bKAGRA
detector \citep{KAGRA_detector}. 


 
The fluctuated noise in the frequency domain,  $n_{sn}(f)$,  
has a Rayleigh distribution with mean value $\braket{n(f)}$. 
Applying an Inverse Fast Fourier Transform,  
the KAGRA noise signal in the time domain, $n_{det}(t_i)$, 
is obtained.


The detector's response for the long wavelength approximation, where
the arm length of the detector is smaller than the reduced wavelength
$\lambda$/2$\pi$, can be described with detector coordinates (latitude
$\lambda$, longitude $L$, angle between East and bisector of the
detector arms $\gamma$, angle between detector arms $\zeta$), GW
source coordinates (right ascension $\alpha$, declination $\delta$,
inclination angle from symmetric axis $\iota$ in 2D numerical
simulation model), local sidereal time of the detector's site
$\phi_t$+$\Omega_r t$, and polarization angle $\Phi$.  These effects
can be written as beam-pattern functions, $F_+$, $F_{\times}$.  Using
an axisymmetric explosion model, the inclination angle between $+Z_w$
axis should be taken into account. The detailed calculation is
done in \cite{beam-pattern}. In case of a gravitational wave emitted
from the light source traveling in the $+Z_w$ direction, the signal
$h$ can be written with two independent waves' polarizations.  That
is, the simulated detector signal is written as follows;
\begin{equation}
 s_{det}(t) = F_{+}(\alpha, \delta, \phi, \theta, \phi_r, t_i)h_{+}(r,\iota, t_i) +  F_{\times}(\alpha, \delta, \phi, \theta, \phi_r, t_i)h_{\times}(r,\iota, t_i) + n_{det}(t_i).
\end{equation}
The KAGRA detector's coordinates are summarized in
Table \ref{tab/det_cor}.


\subsubsection{Calculation of Signal to Noise Ratio}\label{sec/gw_snr}

To evaluate signal and noise power from obtained $s_{det}(t)$, the
Excess Power Filter \citep{excess_power} and Short Time Fourier
Transform (STFT) are used.  The Excess Power Filter extracts signal
power with given time [$t_s$,$t_s$+$\Delta t$], where $t_s$ and
$\Delta t$ are a start time and a time duration of STFT data,
respectively, in a frequency band [$f_s$, $f_s$+$\Delta f$].  In this
paper, the duration time $\Delta t$ is fixed at 31.25 ms and the
frequency band is [40,1000] Hz. We remove the peak frequency of
thermal suspension noise when extracting signal power.

The simulated signal is whitened via a whitening filter to flatten
the noise spectrum in the frequency domain. The whitened signal
${\tilde S_w(f)}$ is calculated by
\begin{equation}
{\tilde S_w(t_s,f)} = \frac{{\tilde S(t_s,f)}}{\braket{{\tilde N(f)}}},
\end{equation}
and ${\tilde S(t_s, f)}$ is calculated by
\begin{equation}
{\tilde S(t_s,f)}=\int_{t_s}^{t_s+\Delta t}s_{det}(t')W(t'-t_s)\exp(-2\pi ift')dt',
\end{equation}
where $\braket{\tilde N(f)}$ is obtained from the running
median \citep{runningmedian} of simulated noise data, and $W(t)$ is a
Hann window function.
 

The signal power $P_s$ and Signal to Noise Ratio (SNR) can be defined
as follows:
\begin{eqnarray}
P_s(t_s) &=& \sqrt{\frac{\int \tilde{S}^*_w(t_s,f) \cdot \tilde{S}_w(t_s,f)df}{\int \braket{\tilde{N}^*_w(f) \cdot \tilde{N}_w(f)}df}}, \\
SNR(t_s) &=& \frac{P_s(t_s) -m}{\sigma}, 
\label{eqn/snr_gw}
\end{eqnarray}
where $m$ and $\sigma$ are the normalized mean and deviation of $P_n
= \sqrt{\int \tilde{N}^*_w(f)\cdot \tilde{N}_w(f) df}$
distribution, respectively. To obtain $m$ and $\sigma$ we first
evaluate the noise behavior, without injecting a supernova signal, and
evaluate $P_n$.  The noise distribution shows $m$=1 and $\sigma$=0.14.

\subsubsection{Extracting Start Time of GW Emission}\label{sec/gw_time}

\begin{figure}[ht]
 \centering
 \includegraphics[width=0.7\textwidth]{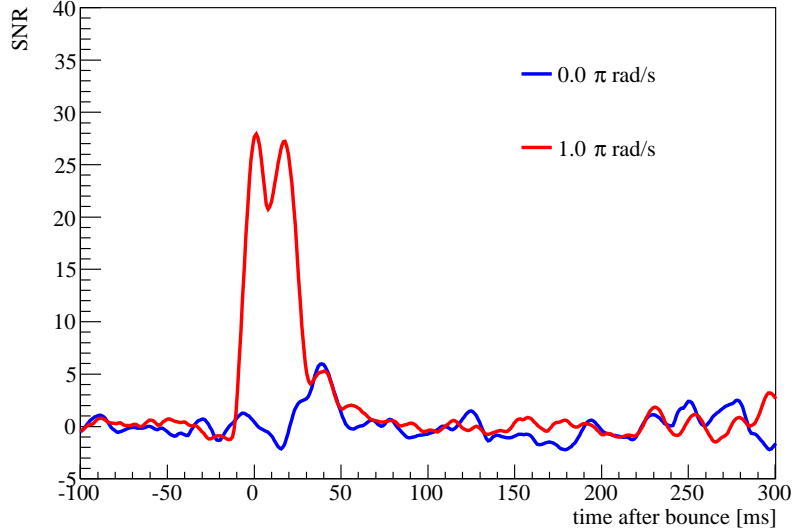}
 \caption{One example of the time variation of obtained SNR for each model.
          Each color shows one progenitor core rotation model,
          0.0$\pi$(blue) 
          and 1.0$\pi$(red) rad s$^{-1}$, respectively.
          The supernova distance is set to 10 kpc and oriented for 
          optimal detection by the KAGRA detector.
         }
 \label{fig/gw_oneshot}
\end{figure}

The center of the timing window which contains the first 
local maximum SNR
is defined as ${\cal T}_{GW}$,
the start time of GW emission.
Figure \ref{fig/gw_oneshot} shows one example of the time variation of
obtained SNR for each rotation model.  The supernova distance is set
to 10 kpc and positioned in the optimal orientation for detection 
by the KAGRA detector.   
In this case, each
supernova simulation gives ${\cal T}_{GW}$ values of 
33 ms (0.0$\pi$ rad s$^{-1}$, blue) 
and -1.0 ms(1.0$\pi$ rad s$^{-1}$, red), 
respectively.
 

We define the detection threshold for GW analysis as 
the first local maximum of the SNR $>$ 8.   
This threshold corresponds to a False Alarm Ratio
of about $\sim$ 10$^{-6}$ per year.

\begin{figure}[ht]
 \begin{minipage}[t]{0.49\textwidth}
  \includegraphics[width=\textwidth]{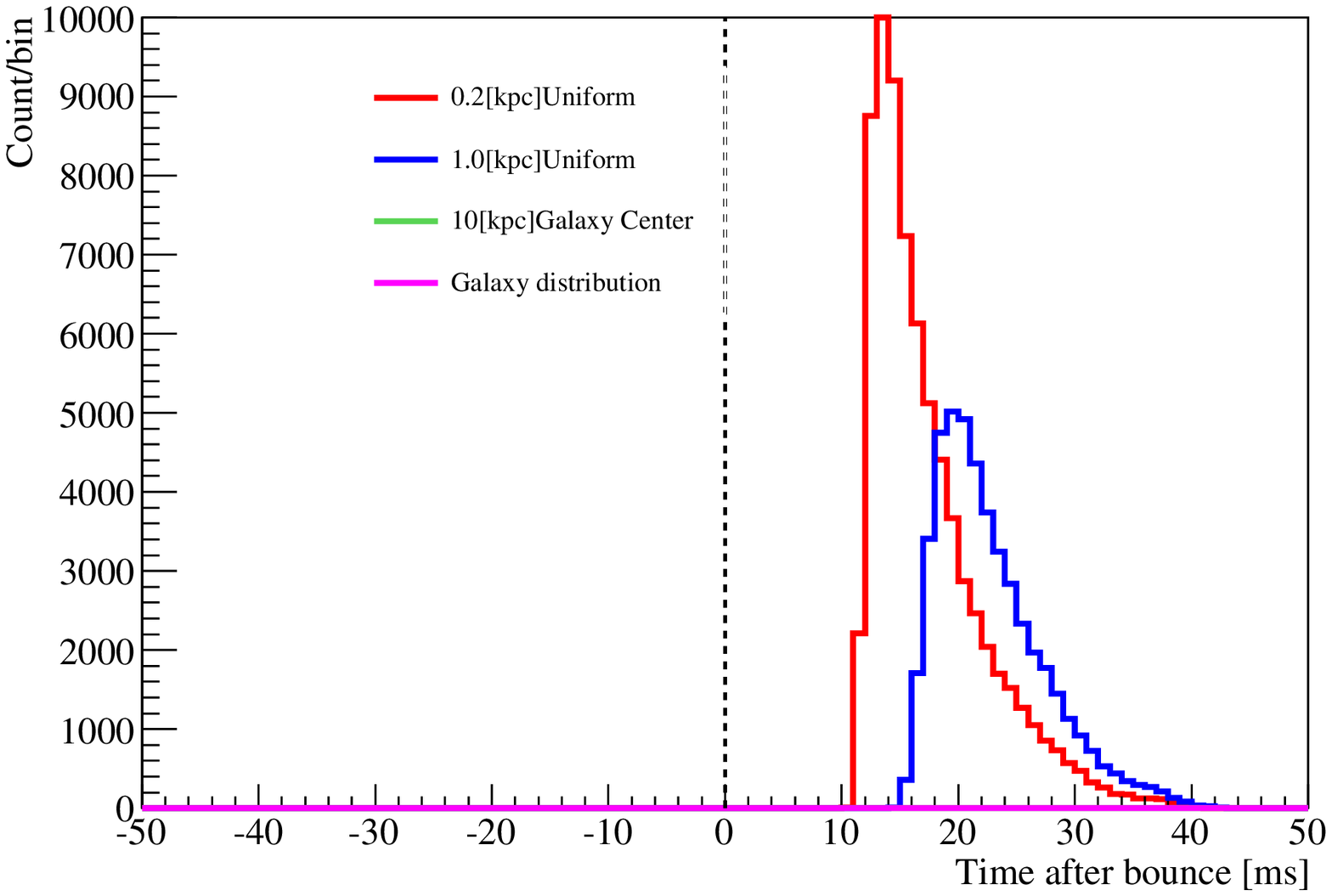}
  \caption{The {\cal T}$_{GW}$ extracted start time distribution 
          of GW emission for each explosion model. 
          The rotation is fixed at 0.0$\pi$(red) rad s$^{-1}$.
          Horizontal axis shows time from core bounce.} 
  \label{fig/gw_time_00}
 \end{minipage}
 \hspace{0.01\textwidth}
 \begin{minipage}[t]{0.49\textwidth}
  \includegraphics[width=\textwidth]{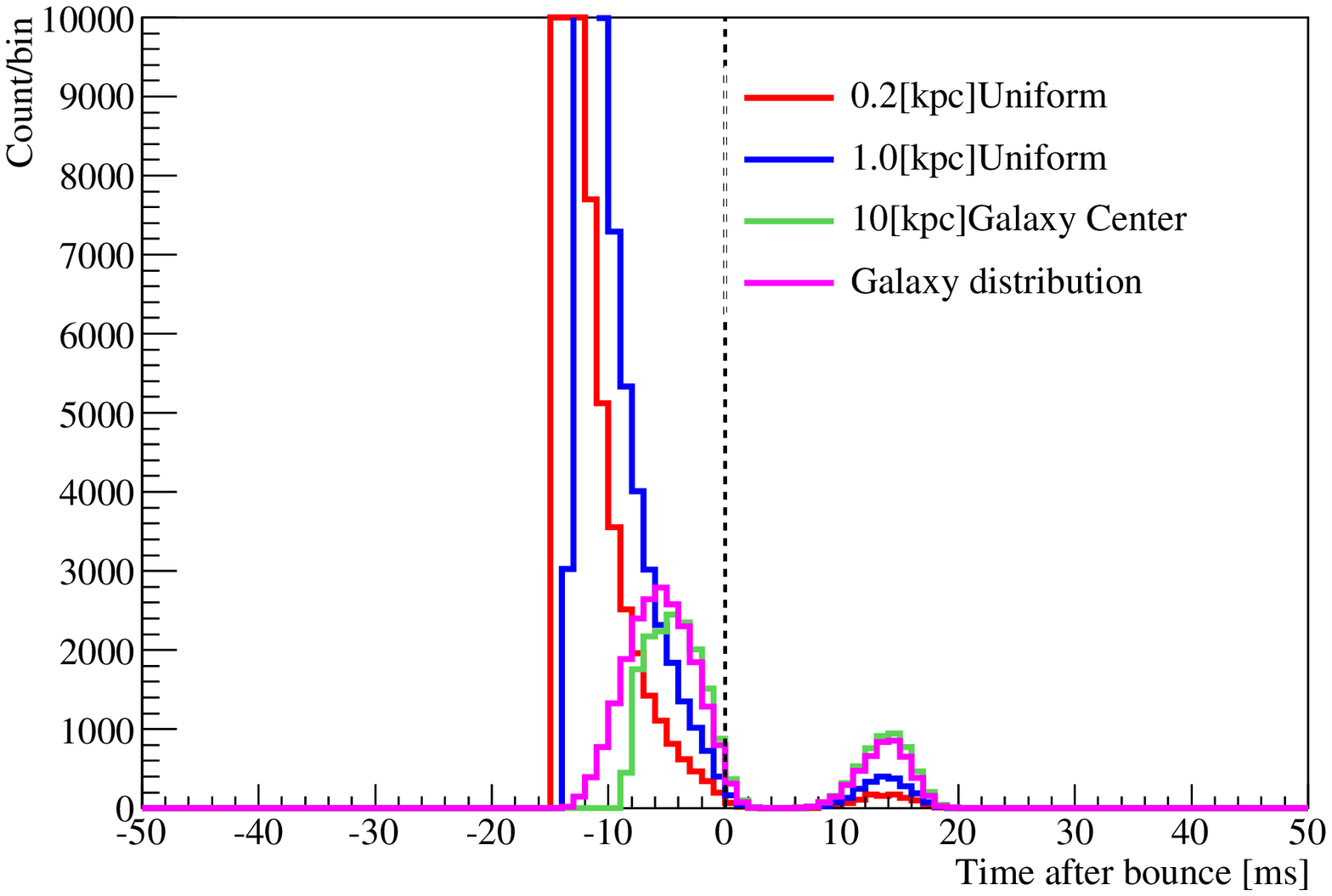}
  \caption{The {\cal T}$_{GW}$ distribution of GW emission for each 
explosion model. 
          The rotation is fixed at 1.0$\pi$(red) rad s$^{-1}$.
          Horizontal axis shows time from core bounce.} 
  \label{fig/gw_time_10}
 \end{minipage}
\end{figure}

Figures \ref{fig/gw_time_00} and \ref{fig/gw_time_10} show
the ${\cal T}_{GW}$ distributions for each explosion rotation model 
for the four supernova scenarios being simulated:
0.2 kpc (red) and 1.0 kpc (blue) uniformly distributed, galactic center 
(green), and galactic distribution (magenta). 
The horizontal axis shows the time after core bounce, when the central
progenitor core density becomes maximum.  When the progenitor core is
strongly rotating, ${\cal T}_{GW}$ is almost the same time as core
bounce, with sharper distributions for closer supernova explosions.
Unfortunately, due to the second peak of $h(t)$ in the 1.0$\pi$ rad
s$^{-1}$ model which can be found around 15 ms in Figure \ref{2-gw},
some simulations show mis-identification of core bounce time.

\subsubsection{Model Dependence}
 
To evaluate the uncertainty of GW epoch extraction we apply the same
analysis to supernova models provided by \cite{dfm}. Of these, we
select the single centrifugal bounce models, which are marked with
crosses in their Table III, because we are interested in fast rotating
models. There are 25 models under this classification.

\begin{figure}[ht]
 \centering
 \includegraphics[width=0.8\textwidth]{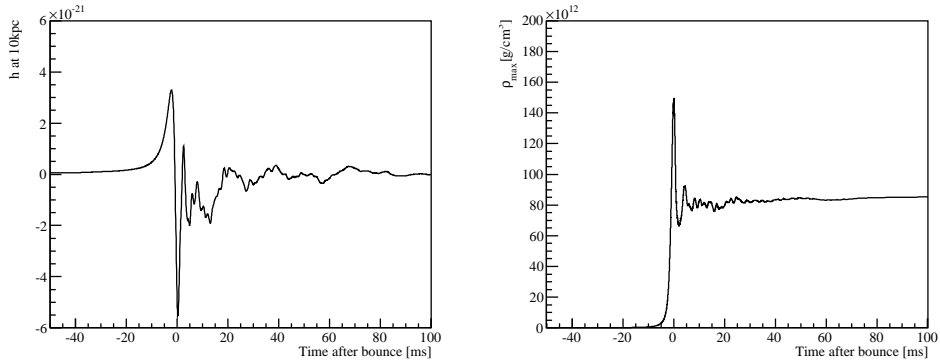}
 \caption{One example of the Dimmelmeier model, e20b-ls in the paper\citep{dfm}. 
          (Left) Time variation of gravitational wave amplitude h 
          as a function of time from core bounce for 10 kpc.
          (Right) Time variation of maximum density $\rho_{max}$.
          }
 \label{fig/gw_dfm_sample}
\end{figure}

 One example of the Dimmelmeier model for gravitational wave amplitude 
$h_{Dim}(t)$ and 
the time evolution of the maximum density $\rho_{Dim,max}(t)$ is 
shown in Fig. \ref{fig/gw_dfm_sample}. 
This model is called 'e20b-ls';  
the progenitor mass is 20 M$_{\odot}$ and the initial state is given by 
stellar evolution simulation. 

\begin{figure}[ht]
\centering
  \includegraphics[width=0.8\textwidth]{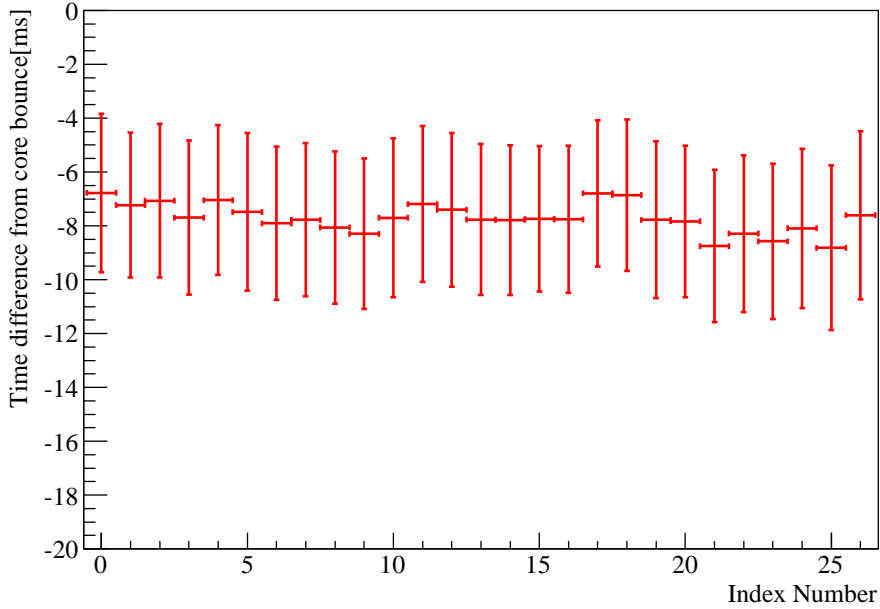}
  \caption{The DFM waveform dependence of the time
          difference from core bounce time for 26 Dimmelmeier models.
          }
  \label{fig/gw_dfm_time}
\end{figure}

Figure \ref{fig/gw_dfm_time} 
shows the extracted time distributions 
when applying the above threshold to the 26 Dimmelmeier models. 
For all such models, the extracted  
times are within about -1$\pm$2 ms of our 1.0 $\pi$ rad/s model. 
This result means that for strong GW models 
we can extract GW emission time with a few ms uncertainty.


\subsection{Neutrino Analysis}
We will now present our method for estimating time variation of the
expected number of neutrino interactions and extraction of
neutronization burst time.

As an aside, we would like to discuss the effect of neutrino mass on
neutrino speed. If the neutrino mass is assumed to be 0.1 meV, derived
from the current direct observational limit, cosmological
limit \citep{wmapa}, and neutrino-less double beta decay experiment
limit \citep{wbeta}, the latency versus light speed is of order 0.1 ms
for an explosion at the center of the galaxy. 
Thus, the mass of neutrinos does not affect the following discussion.

As a further simplification to the analysis, we do not take neutrino
oscillation effects into account, and only consider the electron
flavor neutrino interactions while neglecting those of the $\mu$, and
$\tau$ flavor neutrinos.

\subsubsection{Expected Number of Interactions}\label{sec/nuanal}

From neutrino luminosity, $L_{\nu}(t)$, and mean energy,
$\braket{E_{\nu}(t)}$ of Fig. \ref{2-neutrino}, we obtained an energy
distribution, $dn/dE(t)$, assuming a Fermi-Dirac distribution
\begin{eqnarray}
\braket{E_{\nu}(t)} = \frac{F_3}{F_2}T(t) \approx 3.15T(t), \nonumber\\
f(E,t) = \frac{E^2}{1+\mbox{exp}(E/T(t))}, \nonumber\\ 
\frac{dN}{dE(t)} = f(E,t) \times \frac{L_{\nu}}{F_3 T^4(4\pi r^2)},
\end{eqnarray}
where $T(t)$ is absolute temperature for given time, 
$r$ is distance from Earth, 
and $F_2$ and $F_3$ is defined as
\begin{eqnarray}
&&F_k = \int_0^{\infty}\frac{x^k}{1+\mbox{exp}(x)}, \nonumber\\ 
&&F_2 \approx 1.803, \nonumber\\ 
&&F_3 \approx 5.683. 
\end{eqnarray}

 To identify the time of the neutronization burst,
the expected event rate must be obtained.
We consider three types of interactions in the EGADS and GADZOOKS! detectors. 
The first one is electron neutrino-electron elastic scattering,
\begin{equation}
\nu_e +e \to \nu_e +e.
\label{eqn/nue}
\end{equation}
This is the primary interaction with which to identify the
neutronization burst.  The cross section of the reaction is described
in \cite{nue}. The second one is electron anti-neutrino
elastic scattering,
\begin{equation}
\bar{\nu_e} +e \to \bar{\nu_e} +e.
\end{equation}
The cross section is also described in \cite{nue}.  Because this
reaction occurs only in neutral current interactions, the cross
section is six times smaller than that of electron scattering
(Eq. \ref{eqn/nue}), but this interaction will be a background of the
neutronization burst search.  The final one is inverse beta decay,
\begin{equation}
\bar{\nu_e} + p \to e^+ + n.
\end{equation}
The cross section is described in \cite{nuebar}. Gadolinium-loaded
water Cherenkov detectors have the ability to separate inverse beta decay
reactions from other reactions via tagging of the follow-on neutron.
Because neutron tagging efficiency is roughly 90$\%$, 10$\%$ of these
interactions remain as background.

There are other events generated by supernova neutrinos in water
Cherenkov detectors, such as 
(i) Neutrino-Oxygen interactions,
(ii)$\mu$, $\tau$ neutrino-electron elastic scattering, and so on. But
the rates of these interaction will be small, so that we do not take
them into account in the following analysis. Also, because the
duration of a neutronization burst is $\sim$10 milliseconds, the detector
background may also be ignored ($\sim 10^{-6}$ events s$^{-1}$).

\begin{figure}[ht]
 \begin{minipage}[t]{0.49\textwidth}
  \includegraphics[width=\textwidth]{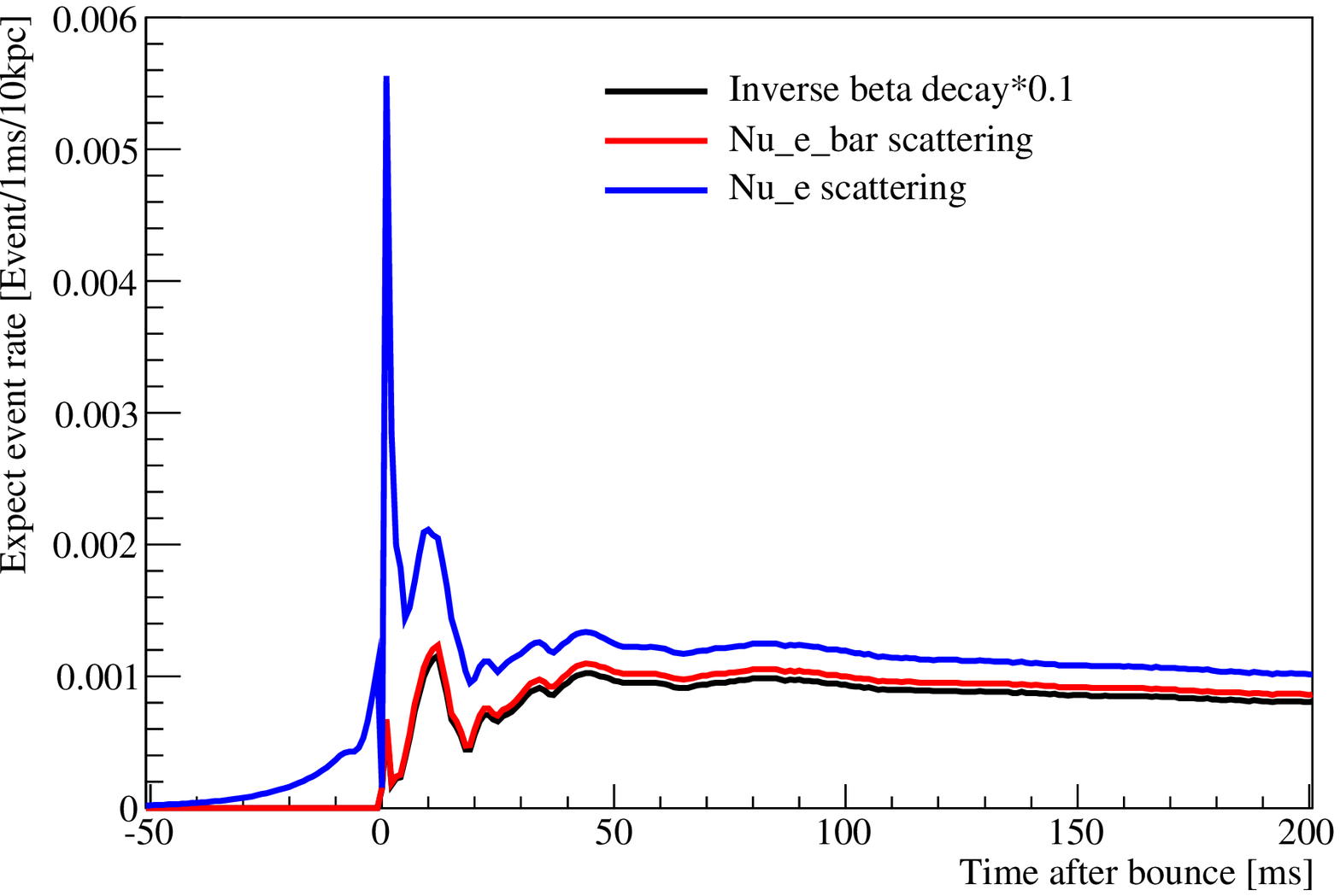}
  \caption{Expected number of interactions in EGADS. 
           Black component shows electron neutrino-electron elastic scattering, 
           blue component shows electron 
           anti-neutrino-electron elastic scattering,
           and red component shows 10$\%$ of inverse beta decay interaction.
           Horizontal axis shows time and vertical 
           axis shows unit of event/1ms/10kpc/100ton.
          }
  \label{fig/exprate}
 \end{minipage}
 \hspace{0.01\textwidth}
 \begin{minipage}[t]{0.49\textwidth}
  \includegraphics[width=\textwidth]{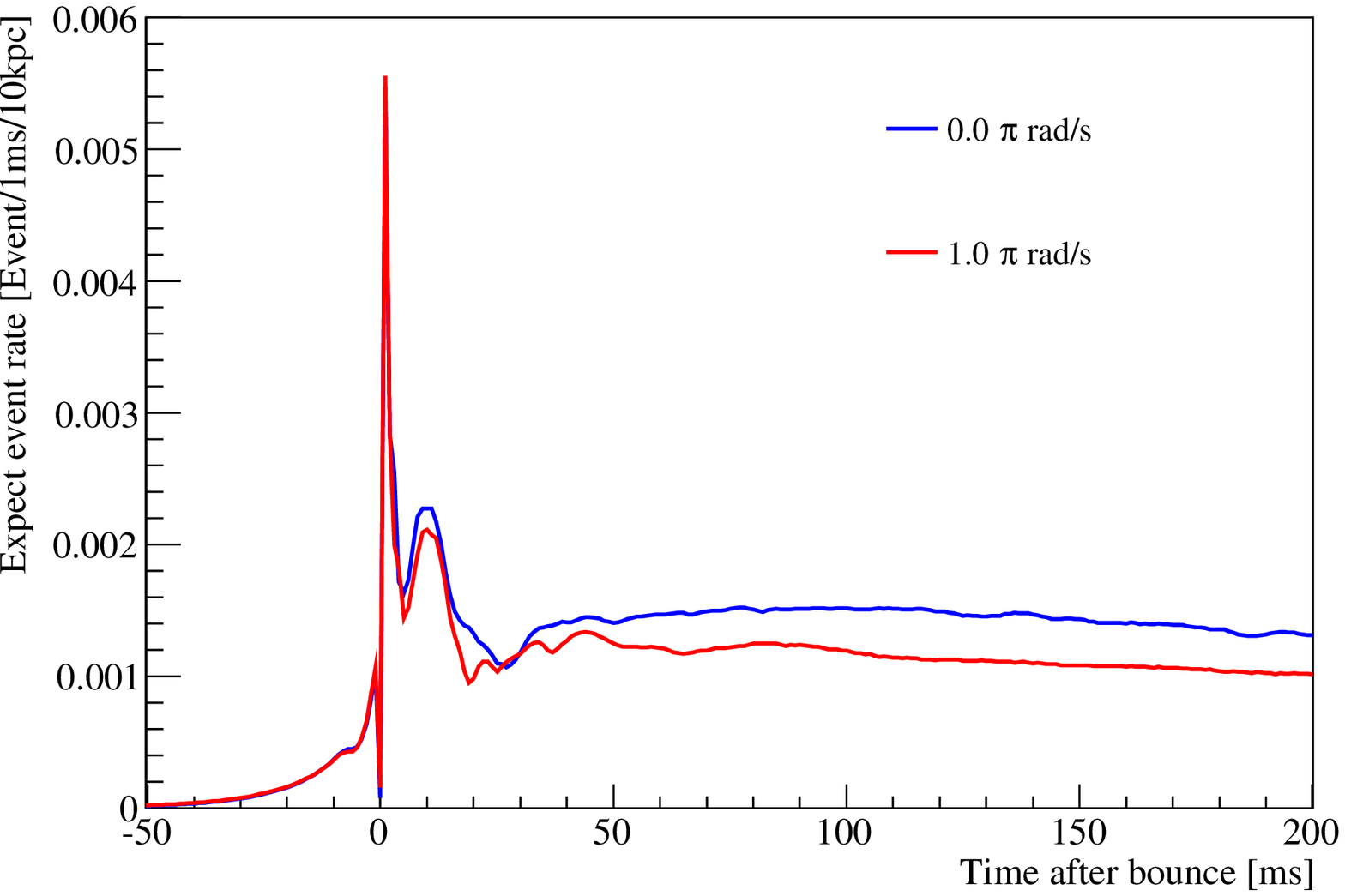}
  \caption{Model dependence of expected number of summed interaction in EGADS. 
          Each color shows one progenitor core rotation model,
          0.0$\pi$(blue)
          and 1.0$\pi$(red) rad s$^{-1}$, respectively.
          Horizontal axis shows time and 
          vertical axis is the sum of expected interactions in EGADS 
          in units of event/1ms/10kpc/100ton. }
  \label{fig/expratemodel}
 \end{minipage}
\end{figure}

Figure \ref{fig/exprate} shows the expected number of these three
interactions in EGADS in the case of a supernova explosion near the
center of our galaxy for the model with $\Omega_0=0.0 \pi$ rad
s$^{-1}$.
 Figure \ref{fig/expratemodel} shows the progenitor core
rotation dependence of expected interaction rates.  
If the progenitor
core is strongly rotating, the gravitational energy is converted to rotation energy, 
in which case the neutrino luminosity becomes smaller as
compared with weaker rotation models.
In the GADZOOKS! detector simulation, the volume of the detector is 225 times 
larger than the EGADS detector.

\begin{figure}[ht]
 \begin{minipage}[t]{0.49\textwidth}
  \includegraphics[width=\textwidth]{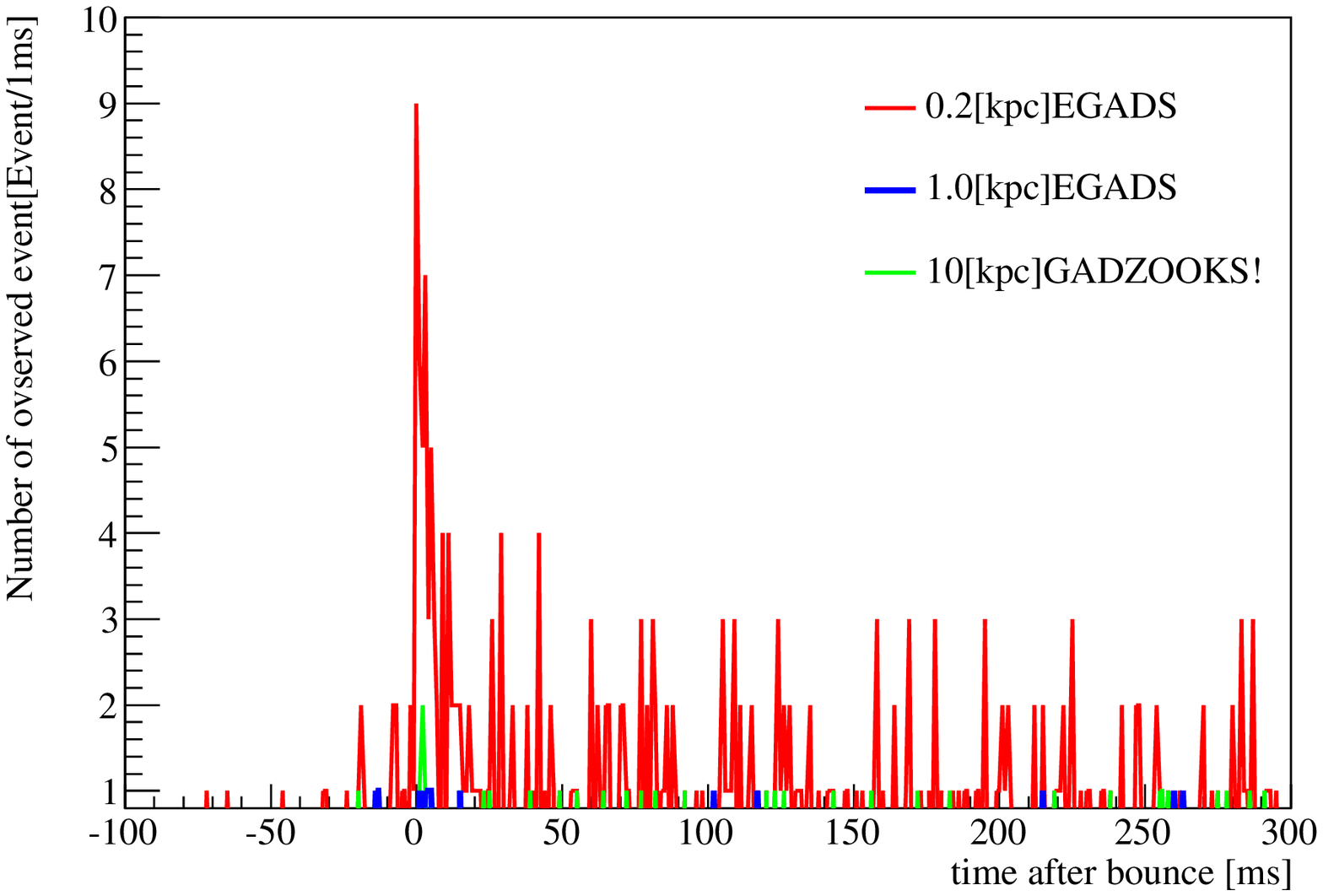}
   \caption{Fluctuation of the number of neutrinos observed by EGADS 
            from distances of 0.2 kpc (red) and 1.0 kpc (blue), and 
            by GADZOOKS! from a distance of 10 kpc (green). 
            The 1.0 $\pi$rad s$^{-1}$ model is used for this figure. 
           }
 \label{fig/nu_oneshot}
\end{minipage}
\hspace{0.01\textwidth}
\begin{minipage}[t]{0.49\textwidth}
 \includegraphics[width=\textwidth]{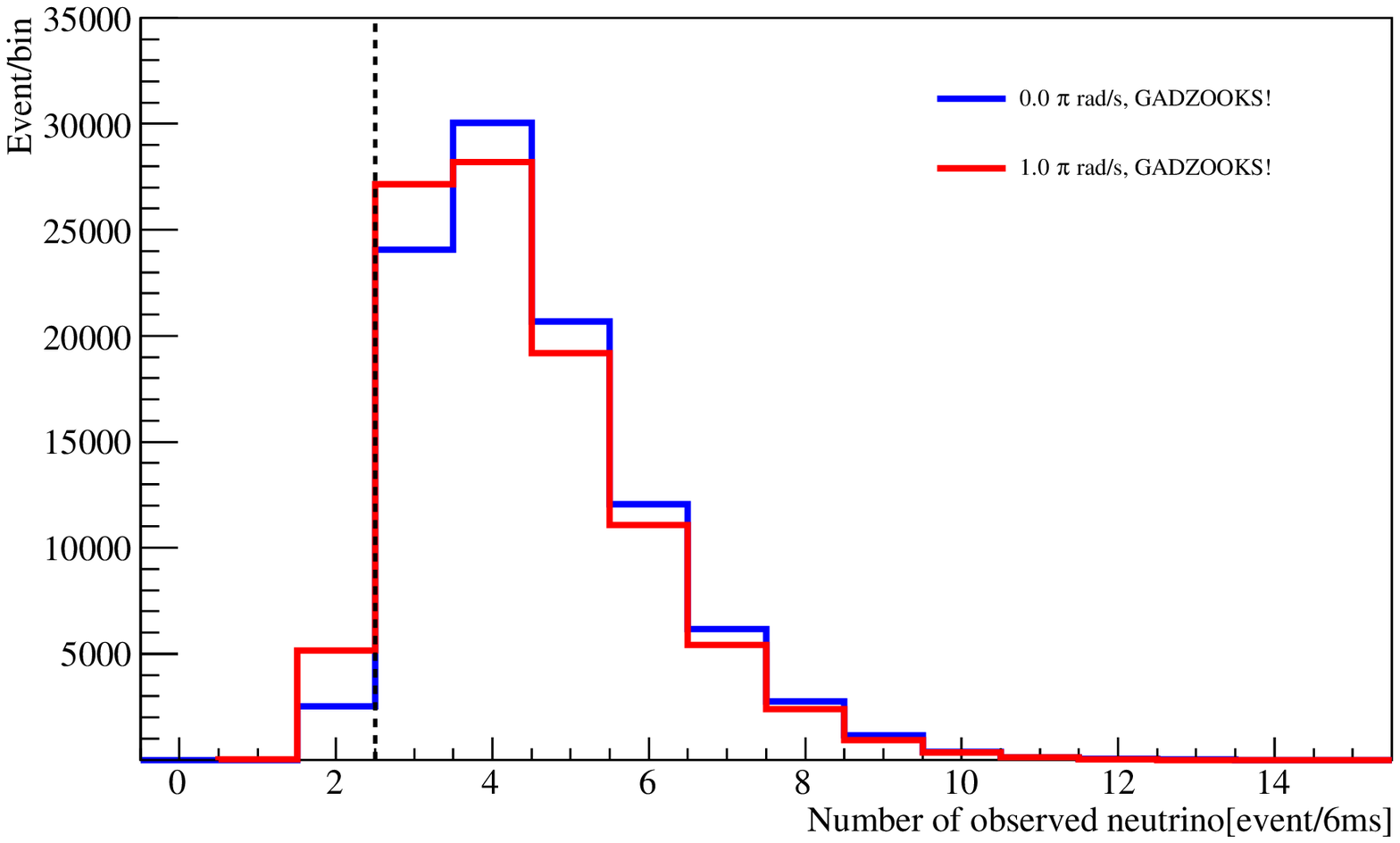}
 \caption{The number of maximum observed neutrinos used as a threshold. 
          The GADZOOKS! detector and a burst at 10 kpc are assumed.
          Colors show the progenitor core rotation models,  
          0.0$\pi$ (blue) and 1.0$\pi$ (red) rad s$^{-1}$, respectively.
          Horizontal axis is number of observed events.
         }
 \label{fig/nu_thres}
 \end{minipage}
\end{figure}

 We employ Poisson statistics to represent the fluctuation of the
observed number of neutrinos within 1 ms intervals.
Figure \ref{fig/nu_oneshot} shows one example of the time fluctuation
of observed neutrinos for for the model with $\Omega_0=1.0\pi$ rad
s$^{-1}$. 
The distances and detectors considered are EGADS for explosions at 
0.2 kpc (red) and 1.0 kpc (blue), and GADZOOKS! for explosions at 
10 kpc (green).

 
\subsubsection{Extracting Neutronization Burst Time}

The neutronization burst time, ${\cal T}_{\nu}$, is obtained via the
following steps:
(i) Open 6 ms sliding window.
(ii) Count the number of observed neutrinos.
(iii) Shift 1 ms and calculate again.
(iv) If maximum observed number of neutrinos exceeds three events, we
define this to be an observation of the neutronization burst.
(v) The center of the time window containing the maximum 
observed number of neutrinos is defined as ${\cal T}_{\nu}$.
If there are multiple candidates for ${\cal T}_{\nu}$, the leftmost 
timing window is defined as ${\cal T}_{\nu}$. Figure
\ref{fig/nu_thres} shows the maximum observed number
of neutrinos for each model.  As already shown in
Fig. \ref{fig/expratemodel}, the model dependence is small.

\begin{figure}[ht]
 \centering
 \includegraphics[width=0.7\textwidth]{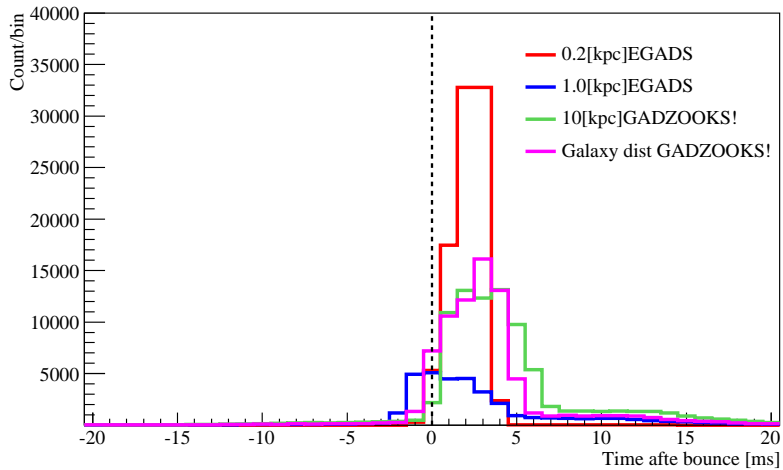}
 \caption{The obtained {\cal T}$_{\nu}$ distribution for 0.2 kpc (red)
          and 1.0 kpc (blue) with EGADS, and for 10 kpc (green) and galactic 
          distribution with GADZOOKS!.}
 \label{fig/nu_time_result}
\end{figure}

Figure \ref{fig/nu_time_result} shows the ${\cal T}_{\nu}$
distributions for the 1.0$\pi$ rad s$^{-1}$ model at 0.2 kpc and 1.0 kpc
as seen by the EGADS detector, and at 10 kpc and distributed 
throughout the galaxy as seen by the GADZOOKS! detector.  
For close supernova explosions, the resulting ${\cal T}_{\nu}$
distribution is quite narrow, and the ${\cal T}_{\nu}$ is estimated as
expected. But as the supernova distance becomes greater, the number of
observed neutrinos per burst becomes smaller, proportional to the usual 
$r^{-2}$ with $r$ being the distance. Therefore, the 
${\cal T}_{\nu}$ distribution fluctuates, and statistical 
uncertainties become larger with increasing distance to the progenitor.


\subsection{Coincidence Analysis}\label{sec/coincidence}
\subsubsection{Definition of Progenitor Core Rotation Estimator}

 By comparing two parameters, 
${\cal T}_{GW}$ and ${\cal T}_{\nu}$, 
we will be able to ascertain the probability that the progenitor's 
core was rotating at the time of collapse.

\begin{figure}[ht]
 \begin{minipage}[t]{0.49\textwidth}
  \includegraphics[width=\textwidth]{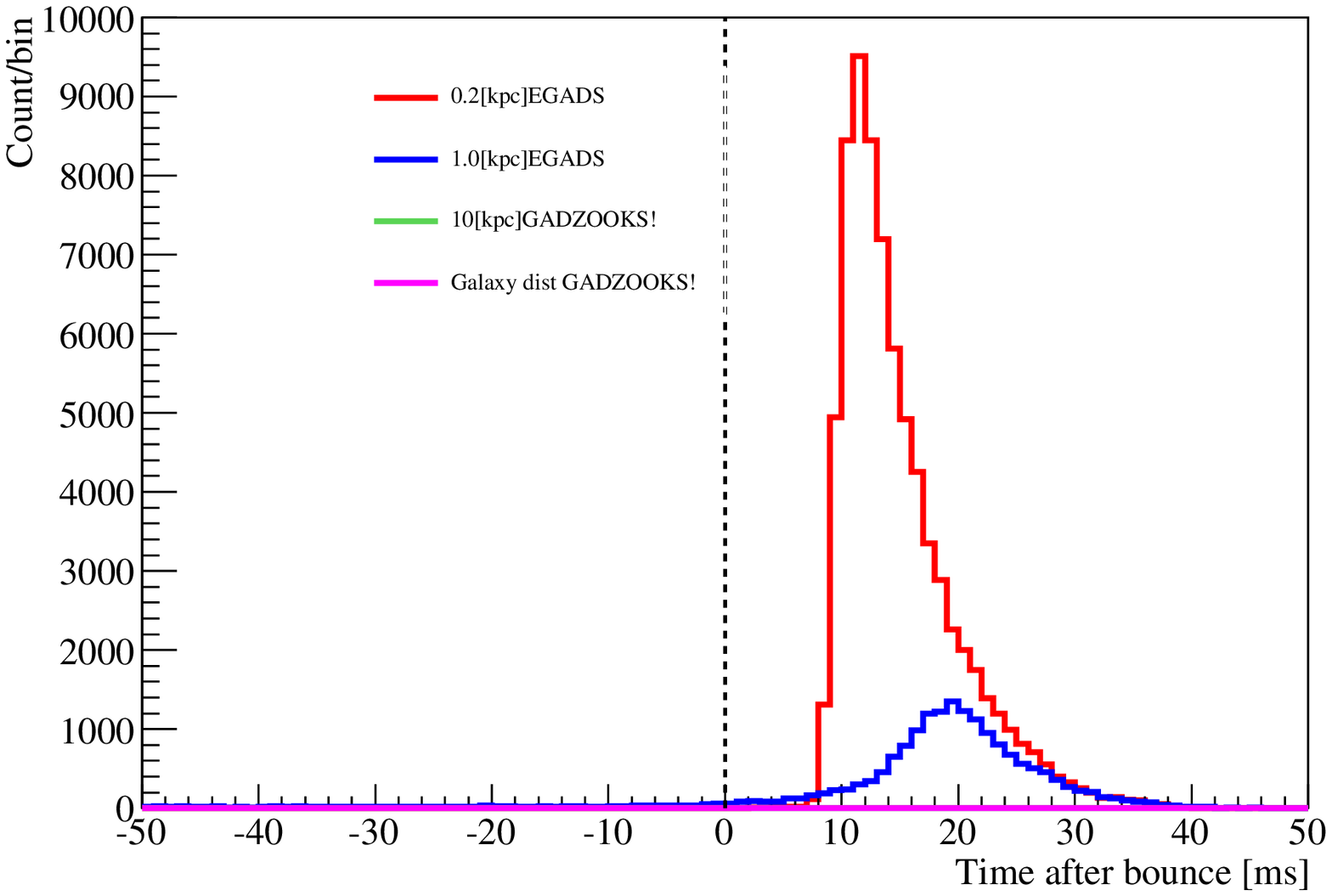}
   \caption{Time comparison between {\cal T}$_{GW}$  
            and {\cal T}$_{\nu}$ 
            distribution for 0.0$\pi$ rad s$^{-1}$ model. 
	    EGADS is used for 0.2 kpc (red) and 1.0 kpc (blue), while
            GADZOOKS! is used for 10 kpc (green) and the galactic 
            distribution (magenta).
            Both GW and neutrino distributions are made
            after applying detection threshold. Horizontal axis shows time from 
            core bounce.}
  \label{fig/comp_time_01kpc_00}
 \end{minipage}
 \hspace{0.01\textwidth}
 \begin{minipage}[t]{0.49\textwidth}
  \includegraphics[width=\textwidth]{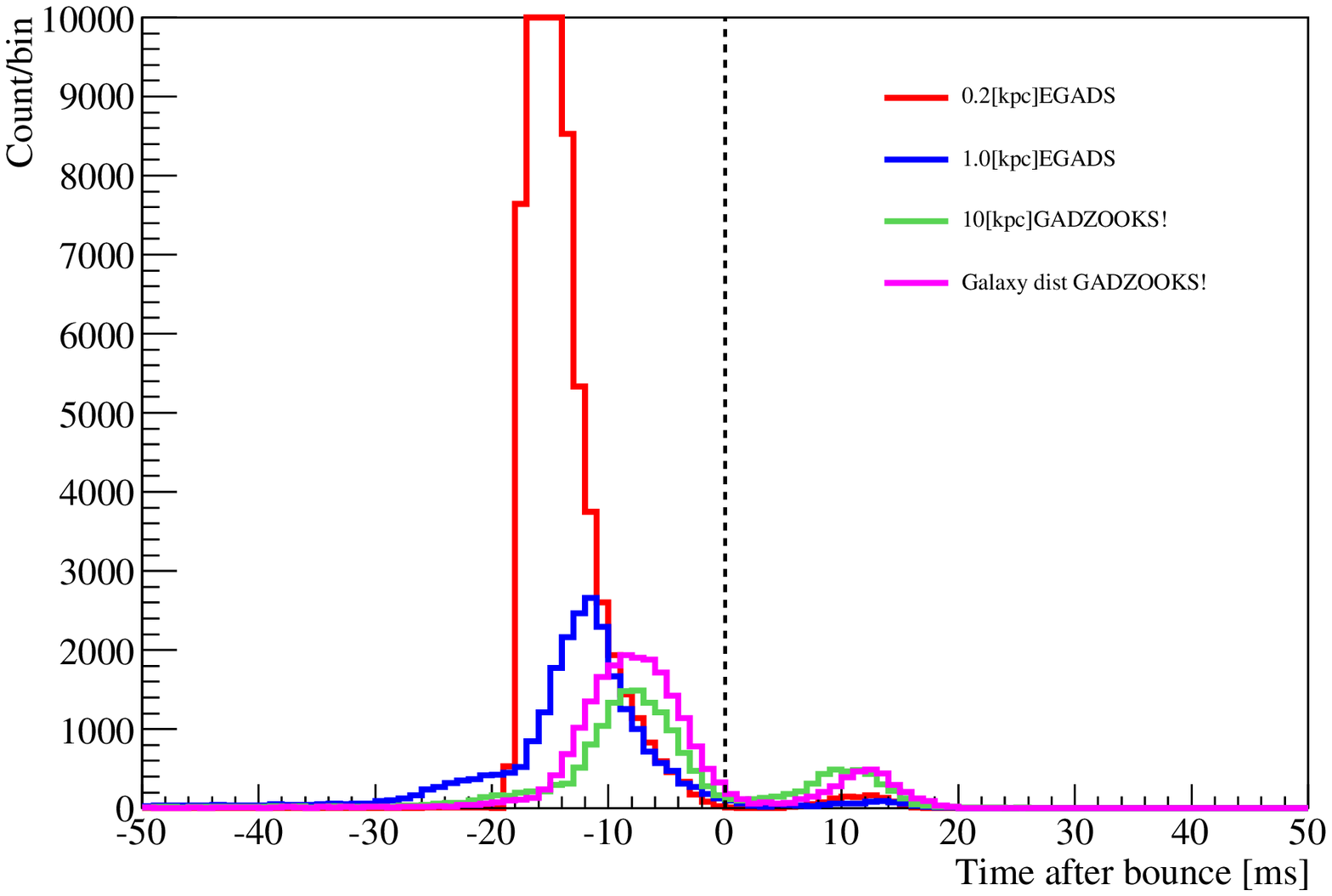}
   \caption{Time comparison between {\cal T}$_{GW}$     
            and {\cal T}$_{\nu}$
            distribution for 1.0$\pi$ rad s$^{-1}$ model.
            EGADS is used for 0.2 kpc (red) and 1.0 kpc (blue), while 
            GADZOOKS! is used for 10 kpc (green) and the galactic 
            distribution (magenta).
            Both GW and neutrino distributions are made
            after applying detection threshold. Horizontal axis shows time from
            core bounce.}
  \label{fig/comp_time_01kpc_10}
 \end{minipage}
\end{figure}

Figures \ref{fig/comp_time_01kpc_00} and \ref{fig/comp_time_01kpc_10} show 
the distributions of ${\cal T}_{GW}$ and ${\cal T}_{\nu}$ 
for the 
0.0$\pi$ rad s$^{-1}$ and 1.0$\pi$ rad s$^{-1}$ models, respectively.
Except for a minor peak in the ${\cal T}_{GW}$ distribution for 1.0$\pi$ rad s$^{-1}$ model, 
a simple comparison between these two times seems to be enough for discussion of 
progenitor core rotation.
 So, the definition of progenitor core rotation is as follows:
we calculate $t_{c}$ which is defined as
\begin{equation}
t_{c} = {\cal T}_{GW} - {\cal T}_{\nu},
\end{equation}
and if $t_{c}<$0, we suppose progenitor core rotation, while for $t_{c}>$0, 
we suppose NO core rotation. 

\subsubsection{Results}

Tables \ref{tab/suwa00} and \ref{tab/suwa10} summarize the $P_r$ values 
for each progenitor core rotation model and scenario, where  
$P_r$ is defined as the estimator for core rotation when both GW and 
neutronization burst signals have been observed.
From these $P_r$ values, we have obtained the following conclusions:\\
 (1) For non-rotating model, (0.0$\pi$ rad s$^{-1}$),  
the $P_r$ value is expected to be close to 0$\%$. 
In the cases with high observed neutrino statistics
the $P_r$ value is indeed almost 0$\%$ as expected. 
But in the case of low neutrino statistics, the $P_r$ value becomes larger 
which leads in turn to reduced determination accuracy of the neutronization 
burst time (${\cal T}_{\nu}$) as shown in Fig. \ref{fig/nu_time_result}.\\
 (2) For the rapidly rotating model (1.0$\pi$ rad s$^{-1}$), 
the $P_r$ value is, as expected, close to 100$\%$. 
But, as shown in Fig. \ref{fig/comp_time_01kpc_10}, there are two peaks due to 
mis-identifying {\cal T}$_{GW}$. This makes for a lower $P_r$ for the more 
distant explosions. Still, even for the galactic cases, the $P_r$ value 
exceeds 70$\%$.

\section{Summary and Discussion}\label{sec/summary}
By using a consistent supernova explosion model emitting both 
GWs and neutrinos, we investigate the progenitor core 
rotation to compare the 
GW emission start time, ${\cal T}_{GW}$, obtained from KAGRA detector, 
and neutronization burst time, ${\cal T}_{\nu}$,  
which is obtained from EGADS and GADZOOKS! detector.
The results show if a nearby supernova is very close(in case of 0.2 kpc), 
we can correctly determine no or slow core rotation about 100$\%$ of the time  
if the progenitor core is indeed not or slowly rotating, and determine the presence of 
core rotation almost 100$\%$ if
the progenitor core is rapidly rotated. 
But we investigate only nearby supernovae using only a single
GW and neutrino detector for this analysis.\\
 
For future studies a coherent or coincidence analysis using multiple GW 
detectors would help to improve the detection efficiencies.
Beyond these potential points for improvement, we currently 
investigate progenitor core rotation with only two parameters, 
${\cal T}_{GW}$ and ${\cal T}_{\nu}$.
Applying multiple classification analysis, the accuracy of $P_r$
is expected to be improved. These things represent our next homework.

Next, we comment on the limitations of our numerical model. 
First, we employed numerical results of two-dimensional (axisymmetric)
Newtonian hydrodynamic simulation of a core-collapse supernova. It is
well known that the hydrodynamic features between 2D and 3D are
different, especially for the cascade direction of turbulent motion
\citep{hank12,couc14,hand14,taki14}. 
The forward cascade (from large-scale to small scale), which is
typical in 3D case, may weaken the large scale prompt convection and
increase the chance being classified as rapidly rotating. This aspect
should be checked by performing 2D and 3D simulations with the same
setups. More important limitations of this study are neglecting
general relativistic (GR) and magnetohydrodynamic (MHD) effects. As
for GR effects, there have been several GR simulations that include
detailed microphysics \citep[e.g.][]{dfm,seki10,ott12,muel13,kuro14},
which implied that GR leads to higher characteristic frequencies of
gravitational wave signals due to more compact protoneutron stars as a
consequence of stronger gravity. As for MHD effects, the magnetic
fields would transfer the angular momentum from inside to outside and
a strong jet is launched if the initial magnetic fields are strong
enough, which would modify the GW
signatures \citep{ober06,sche10,taki11,sawa13}. However, since MHD
effects do not have a large impact around the bounce, our conclusion
will not be affected even if we include magnetic fields in our
simulation for future projects.
Secondly, in this paper we employed one progenitor model, 
s11.2 of \citet{woos02} because it is well studied by a number of 
previous works. It should be noted that the prompt convection depends 
on the structure of the progenitor model. Therefore, to assess the 
robustness of the results obtained in this study we need a more systematic 
study using multiple progenitor models. In addition, we employed one 
specific angular velocity distribution. More study of the dependence 
on the rotation profile is also needed \citep[but see][]{abdi13}.
Thirdly, we employed simple weak interactions in solving
Boltzmann equation of neutrinos, which may affect the convection
driven by unstable configuration of lepton fraction
\citep[e.g.,]{lent12}. Simulations with more detailed microphysics are
 needed.
Finally, we did not take into account heavier leptonic neutrinos and
neutrino oscillation effects.  As described in
Sec. \ref{sec/nuanal}, the elastic scattering cross sections for
heavier leptonic neutrinos are about six times smaller than electron
flavor neutrinos. 
If generated electron neutrinos are converted to 
other heavier flavor neutrinos, 
the height of the neutronization burst would be smaller than 
the present estimation and the detection efficiency of the 
neutronization burst would also be smaller. 
The conversion ratio depends on the neutrino mass hierarchy 
(normal or inverted), 
neutrino oscillation parameters 
(mixing angles and mass differences), 
relations between matter and self-interaction effects in the progenitor, 
and whether the neutrinos first pass through the Earth or not 
\citep{nuosc1, nuosc2}before being detected. 
In particular, the neutrino mass hierarchy is crucially important. 
In the normal hierarchy case, 
almost all of the electron neutrinos generated in the inner core 
will convert to heavier flavor neutrinos via the matter effect 
(neglecting self-interaction effects), 
and it would be difficult to apply the techniques used for this study. 
Even in the inverted hierarchy case, 
some electron neutrinos will be converted to heavier flavor neutrinos; 
this effect would not be negligible.
This issue will
also be investigated in a forthcoming paper.

\vspace*{1cm}
\centerline{\bf Acknowledgments}
\vspace*{0.5cm}
This work was supported by the MEXT Grant-in-Aid for Scientific Research
on Innovative Areas ``New Developments in Astrophysics Through
Multi-Messenger Observations of Gravitational Wave Sources''
(Nos. 24103004, 24103005, 24103006, 25103511), JSPS postdoctoral
fellowships for research abroad, MEXT SPIRE, and JICFuS. Numerical
computations in this study were in part carried on XC30 at CfCA in
NAOJ and SR16000 at YITP in Kyoto University.

\bibliography{reference.bib}

\begin{table}[h]
  \begin{center}
  \centering
  \caption{KAGRA detector coordinates.}
  \begin{tabular}{|c|c|}
    \hline
    parameter & value \\  
\hline \hline
    latitude of KAGRA detector $\lambda$ & +36.41$^{\circ}$ \\ \hline
    longitude of KAGRA detector$L$ & +137.3$^{\circ}$ \\  \hline
    angle between East and bisector of the detector arms$\gamma$ & +75.0$^{\circ}$ \\ \hline
    angle between detector arms $\zeta$ & 90.0$^{\circ}$ \\ \hline
  \end{tabular}
\label{tab/det_cor}     
  \end{center}
\end{table}

\begin{table}[h]
 \centering
 \caption{The various scenarios' GW detection efficiencies (GW eff.),
   neutronization neutrino detection efficiencies (neutrino eff.)  for
   (1) EGADS or (2) GADZOOKS!, their product, and the $P_r$ value for
   the 0.0$\pi$ rad/s model.}
  \begin{tabular}{|c| c| c| c| c|}
\hline
    Scenario & GW eff.[$\%$] & neutrino eff.[$\%$] & detection eff.[$\%$] & $P_r$[$\%$] \\ 
\hline \hline
0.2kpc, uniform  & 74.8 & 100.0(1) & 74.8  & 0.0  \\  \hline
1.0kpc, uniform  & 46.5 & 46.8(1)  & 21.9  & 20.8  \\ \hline
Galactic Center  & 0.0  & 97.5(2)  & 0.0   & ---  \\ \hline
Galaxy Dist.     & 1.5  & 84.6(2)  & 1.5   & 0.2  \\ \hline
  \end{tabular}
  \label{tab/suwa00}
\end{table}

\begin{table}[h]
 \centering
 \caption{The various scenarios' GW detection efficiencies(GW eff.), 
neutronization neutrino detection efficiencies(neutrino eff.) for (1) EGADS or (2) 
GADZOOKS!, their product, and the $P_r$ value for the 1.0$\pi$ rad/s model.}
  \begin{tabular}{|c| c| c| c| c|}
\hline
    Scenario & GW eff.[$\%$] & neutrino eff.[$\%$] & detection eff.[$\%$] & $P_r$[$\%$] \\ 
\hline \hline
0.2kpc, uniform  & 88.0 & 100.0(1) & 88.0  & 98.4  \\  \hline
1.0kpc, uniform  & 73.6 & 40.2(1)  & 29.5  & 80.0  \\ \hline
Galactic Center  & 21.5 & 94.8(2)  & 20.4  & 75.3  \\ \hline
Galaxy Dist.     & 26.7 & 81.7(2)  & 24.7  & 76.2  \\ \hline
  \end{tabular}
  \label{tab/suwa10}
\end{table}

\end{document}